\documentclass[aps,prb,twocolumn,superscriptaddress,floatfix,nofootinbib]{revtex4-1}

\usepackage{graphicx}
\usepackage{amsmath,amsfonts,amssymb}
\usepackage{color}
\usepackage{subfigure}
\usepackage{physics}
\usepackage{dsfont}
\usepackage{hyperref}
\usepackage{url}
\usepackage{leftidx}
\usepackage{textcomp}
\usepackage{gensymb}

\usepackage[utf8]{inputenc}
\usepackage[dvipsnames]{xcolor}
\usepackage{multirow}
\usepackage{comment}
\usepackage{pifont}

\usepackage{colortbl}

\newcommand{\angstrom}{\mbox{\normalfont\AA} }

\usepackage{xr}
\makeatletter
\newcommand*{\addFileDependency}[1]{
  \typeout{(#1)}
  \@addtofilelist{#1}
  \IfFileExists{#1}{}{\typeout{No file #1.}}
}
\makeatother

\newcommand*{\myexternaldocument}[1]{%
    \externaldocument{#1}%
    \addFileDependency{#1.tex}%
    \addFileDependency{#1.aux}%
}

\myexternaldocument{SI_tidy}

\begin{document}

\title{Microscopic Theory of  Magnetic Disorder-Induced Decoherence in Superconducting Nb Films}

\author{Evan Sheridan}
\thanks{These authors contributed equally}
\affiliation{Materials Science Division, Lawrence Berkeley National Laboratory, Berkeley, CA 94720, USA}
\affiliation{Molecular Foundry, Lawrence Berkeley National Laboratory, Berkeley, CA 94720, USA}
\affiliation{King's College London, Theory and Simulation of Condensed Matter, WC2R 2LS London, UK}

\author{Thomas F. Harrelson}
\thanks{These authors contributed equally}
\affiliation{Materials Science Division, Lawrence Berkeley National Laboratory, Berkeley, CA 94720, USA}
\affiliation{Molecular Foundry, Lawrence Berkeley National Laboratory, Berkeley, CA 94720, USA}

\author{Eric Sivonxay}
\affiliation{Energy Technologies Area, Lawrence Berkeley National Laboratory, Berkeley, CA 94720, USA}
\affiliation{Department of Materials Science and Engineering, University of California, Berkeley, Berkeley, CA 94704, USA}

\author{Kristin A. Persson}
\affiliation{Energy Technologies Area, Lawrence Berkeley National Laboratory, Berkeley, CA 94720, USA}
\affiliation{Department of Materials Science and Engineering, University of California, Berkeley, Berkeley, CA 94704, USA}

\author{M. Virginia P. Alto\'{e}}
\affiliation{Molecular Foundry, Lawrence Berkeley National Laboratory, Berkeley, CA 94720, USA}

\author{Irfan Siddiqi}
\affiliation{Department of Physics, University of California, Berkeley, California 94720, USA}
\affiliation{Center for Quantum Coherent Science, University of California, Berkeley, California 94720, USA}

\author{D. Frank Ogletree}
\affiliation{Molecular Foundry, Lawrence Berkeley National Laboratory, Berkeley, CA 94720, USA}

\author{David I. Santiago}
\affiliation{Center for Quantum Coherent Science, University of California, Berkeley, California 94720, USA}

\author{Sin\'{e}ad M. Griffin}
\affiliation{Materials Science Division, Lawrence Berkeley National Laboratory, Berkeley, CA 94720, USA}
\affiliation{Molecular Foundry, Lawrence Berkeley National Laboratory, Berkeley, CA 94720, USA}

\begin{abstract}
The performance of superconducting qubits is orders of magnitude below what is expected from theoretical estimates based on the loss tangents of the constituent bulk materials. This has been attributed to the presence of uncontrolled surface oxides formed during fabrication which can introduce defects and impurities that create decoherence channels.  Here, we develop an \textit{ab initio} Shiba theory to investigate the microscopic origin of magnetic-induced decoherence in niobium thin film superconductors and the formation of native oxides. Our \textit{ab initio} calculations encompass the roles of structural disorder, stoichiometry, and strain on the formation of decoherence-inducing local spin moments. With parameters derived from these first-principles calculations we develop an effective quasi-classical model of magnetic-induced losses in the superconductor. We identify \textit{d}-channel losses (associated with oxygen vacancies) as especially parasitic, resulting in a residual zero temperature surface impedance. This work provides a route to connecting  atomic scale properties of superconducting materials and macroscopic decoherence channels affecting quantum systems.

\end{abstract}

\maketitle

\section{Introduction}


Recent breakthroughs in superconducting qubits, following decades-long improvements in fabrication and design, have established superconductors as a leading platform for scalable fault-tolerant quantum computing\cite{arute_nat19, wallraff_review_20, oliver_aprev19, muller_arxiv19, zmuidzinas_annrevcmp12, murch_nature13a, murch_nature13b, macklin_sci15}. Despite this progress, decoherence in superconducting materials remains a major obstacle for future progress in quantum information systems\cite{muller_arxiv19, oliver_aprev19}, and in other applications including particle detection and quantum sensors in high-energy physics~\cite{zmuidzinas_annrevcmp12}. Central to diagnosing and mitigating the macroscopic decoherence channels in superconducting qubits is an understanding of the behaviour of the constituent materials at an atomic scale, especially considering inevitable inhomogeneities such as defects, structural disorder, and interfaces. 


Decoherence channels in superconducting materials are broadly classified into charge noise, spin impurity magnetic flux noise, and non-magnetic quasiparticle excitations~\cite{oliver_aprev19}. Magnetic flux noise arises through interactions between Cooper pairs and magnetic impurities mediated by an electromagnetic field, and have been suggested to contribute to \textit{1/f} noise~\cite{clarke_apl87, clarke_prl07, Weissman_1988, Pershoguba_et_al:2015}. While BCS superconductors are robust against non-magnetic impurities\cite{anderson_prl59}, over a critical magnetic impurity density a gapless superconductor emerges\cite{gorkov_60}. Scattering on classical spins in s-wave superconductors can cause quasiparticle bound states to appear deep within in the gap, forming so-called Yu-Rusinov-Shiba (YSR) states ~\cite{yu_65,shiba68,rusinov_jetp69}. Time-reversal symmetry breaking induced by magnetic impurities results in pair breaking that suppresses superconductivity. Recently, Pellin et al.~\cite{pellin12} proposed that magnetic impurities in s-wave superconductors cause the surface impedance to plateau with decreasing temperature, such that the decoherence due to magnetic impurities cannot be made arbitrarily small by lowering the temperature. They also show that the magnetic scattering rate required to suppress superconductivity is three orders of magnitude less than the nonmagnetic scattering rate~\cite{pellin12}, emphasizing the need for microscopic models incorporating atomic and chemical specificity of magnetic-induced decoherence in superconductors. 


Previous investigations of local defects in superconducting Al films  identified unpaired spins at the air interface of the superconducting film from the adsorption of molecular oxygen, or the formation of a metallic oxide layer~\cite{degraaf_prl17, degraaf_natcomm18, choi_prl09}. Electron spin resonance found free spins on the interface of superconducting films;  subsequent processing steps reduced the impurity concentration and improved the system's coherence~\cite{degraaf_natcomm18}. Applying a DC-magnetic field in both the in-plane and out-of-plane direction of a NbN coplanar waveguide resonators during operation alters its quality factor -- the authors suggest this is due to the presence of paramagnetic impurities caused by dangling bonds on its surface\cite{zollitsch_aip19}. \textit{Ab initio} calculations suggest that adsorbed molecular oxygen creates a paramagnetic species on aluminium-based superconductors~\cite{adelstein_aip17}, contributing to \textit{1/f} noise consistent with experiments~\cite{wang_prl15}. Similar calculations also find that dangling hydroxyl groups on the surface of Al$_2$O$_3$ can generate a local spins~\cite{lee_prl14}.


In Nb-based superconducting qubits, atomic-scale characterisation with X-ray photoelectron spectroscopy and resonant inelastic X-ray scattering have found that a Nb-oxide layer readily forms at the interface~\cite{halbritter_87, premkumar_arxiv20, altoe_arxiv20}. Magnetic and charge defects in this Nb-oxide layer have been suggested as a significant source of dissipation~\cite{halbritter_87, proslier_ieee_11, pellin12}. Recently, Alto\'{e} et al.~\cite{altoe_arxiv20} found that the SiO$_x$ primarily contributes to the two-level system (TLS) losses, while the Nb oxide layer dominates non-TLS losses in a coplanar waveguide resonator, both of which can be mitigated through surface treatments.


In this work we explore the atomistic origins of magnetic-induced dissipation in superconducting Nb thin films. We develop an \textit{ab initio} Shiba theory for surface impedance that incorporates the realistic atomic structure of  surface oxides to describe the origins of dissipative loss channels. We describe magnetic impurities inside the oxide layer on the surfaces of Nb-based superconducting devices using density functional theory (DFT) to parameterize an Yu-Shiba-Rusinov theory of magnetic impurities for BCS superconductivity. We consider the range of disorder and off-stoichiometry found in the surface oxide by considering crystalline and amorphous phases of Nb-oxides (NbO, NbO$_2$, and Nb$_2$O$_5$) with and without defects, disorder, and strain. From these theoretical results, we suggest fabrication considerations to mitigate the most destructive magnetic decoherence channels in Nb-based superconductors.

\section{Properties of Niobium Oxide Thin Films and Crystal Phases}

\begin{figure*}
    \centering
    \includegraphics[width=1.0\linewidth]{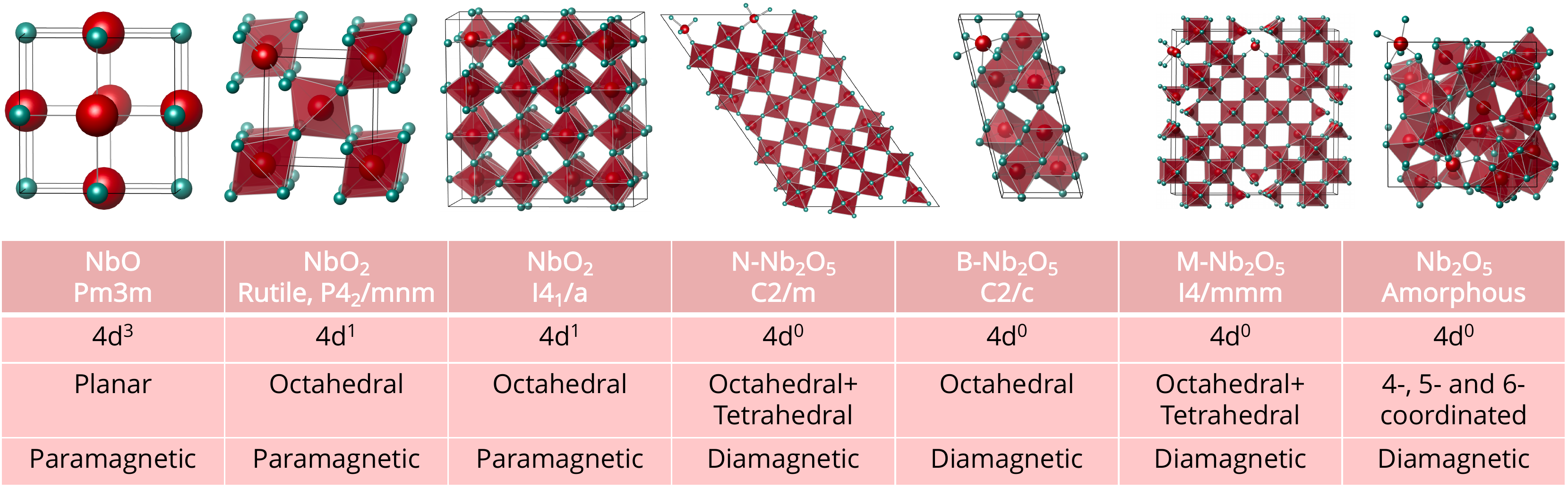}
    \caption{Crystal structures of the Nb oxides considered in this work with their nominal valence configuration on the Nb atoms, the Nb coordination environment(s) and the resulting magnetic order in the stoichiometric case.}
    \label{fig:crystalline_fig_1}
\end{figure*}

Niobium oxides have been studied for potential optoelectronic applications yielding rich insights into the relevant phases and structures~\cite{stormer_09, nico_jpcc_11, nico_pms_16}, in addition to exotic emergent phenomena including metal-insulator transitions\cite{Eyert:2002}. In this work we focus on NbO, NbO$_2$, and Nb$_2$O$_5$~\cite{nico_jpcc_11}(Figure \ref{fig:crystalline_fig_1}), which were identified as the dominant stoichiometries of the Nb-oxides formed on Nb resonators from elemental STEM EELS mapping in Ref.~\cite{altoe_arxiv20}. Transmission electron microscopy measurements from the same work also suggests primarily amorphous Nb$_2$O$_5$, which has been further confirmed with fluctuation electron microscopy measurements and is consistent with X-ray Absorption Spectroscopy performed on the same films\cite{Structure_Paper}. Informed by these measurements, we consider both crystalline and amorphous phases of Nb oxides to calculate the role of structural disorder, defects and strain on the electronic and magnetic properties from first principles. 

NbO adopts a face-centered cubic structure derived from the rocksalt structure but with ordered defects resulting in square-planar coordination~\cite{Burdett_et_al:1984,bowman_66}.  NbO$_2$ exists in either the rutile (metallic) or tetragonal (semiconducting) phase with a small bandgap of 0.5 eV~\cite{nico_pms_16}, both of which comprise of octahedrally coordinated Nb forming edge- and corner-sharing networks. Nb$_2$O$_5$ can be formed by oxidizing NbO, and is polymorphic with as many as eight different phases~\cite{nowak_chemrev_99, nico_jpcc_11, nico_pms_16}, mainly formed from NbO$_{6}$ octahedra. The range of crystalline polymorphs are labelled by letters corresponding to the temperature at which they were obtained including the TT- or T-phase at low temperatures, the M-,  B and N-phases at intermediate temperatures, and the H-phase at higher temperatures. Other phases are stabilized through different thermal processing steps~\cite{nowak_chemrev_99}. As expected from a fully-occupied O-\textit{p} manifold and an empty Nb-\textit{d} manifold, the Nb$_2$O$_5$ polymorphs are wide bandgap semiconductors with gaps varying between 3.4-4.0~eV~\cite{nico_pms_16}. 

The variety of point- and edge-sharing octahedra in Nb pentoxide is a direct result of the versatility of the nominal $d^{0}$ oxidation state and their corresponding physical features~\cite{nowak_chemrev_99}. N-Nb$_2$O$_5$ is described by 4x4 blocks of point-sharing octahdra, where the blocks are connected by edge-sharing octahedra. Tetragonal M-Nb$_2$O$_5$ can similarly be described as a set of of 4x4 blocks, but consists of a mix of point-sharing octahdra and tetrahedra, where adjacent blocks are connected by edge-sharing octahedra. Monoclinic B-Nb$_2$O$_5$ possesses a TiO$_2$(B)-like structure and is described by a set 2x2 blocks of edge-sharing octahedrally distorted NbO$_6$ polyhedra connected by point-sharing octahedra. The pseudo-hexagonal TT-phase of Nb$_2$O$_5$ has also been identified as a metastable variant of the T-phase, with magnetic susceptibility measurements suggesting that O vacancies contribute to a paramagnetic spin density of $2.1-9.9\times 10^{21}$~moments per mole~\cite{herval_15}. In some cases, O defects become ordered resulting in several stable phases that also form magnetic moments~\cite{cava_nature91, waldron_jacs01, fang_jmcc15}. In fact, the slightly O deficient Nb$_{12}$O$_{29}$ is antiferromagnetically ordered\cite{cava_nature91}. 


\section{Methods}

We calculated the electronic structure and magnetic properties of the  Nb oxides using DFT as implemented in the Vienna Ab initio Simulation Package (VASP)~\cite{kresse93}. Our full DFT calculations details are given in the SI; here we justify our  choice of exchange-correlation functional. We apply a Hubbard-U of 4 eV to the Nb-\textit{d} states to account for the localization of these transition-metal orbitals, which is consistent with previous DFT+U studies on Nb-oxides\cite{kocer_prb19}. We further test the sensitivity of our calculated magnetic moments to this choice of \text{U} (see SI \ref{fig:u_study}), finding a very slight increase in the calculated projected magnetic moments as a result of the increased localization of the \textit{d} orbitals. Finally, we calculated the bandgap of B-Nb$_{2}$O$_{2}$ to be 2.89 eV (Figure~\ref{fig:SI_HSE}), underestimating the experimental value of ~3.8-3.9 eV \cite{pinto_jpca_17}, as expected for semilocal DFT exchange-correlation functionals\cite{bandgap_problem}. Improvements to the treatment of electron delocalization using a screened hybrid functional can give closer results to experiment\cite{hse06}; in our case  HSE06 opens the bandgap to 3.86 eV. However, we find that the qualitative features of the calculated density of states (DOS) do not change between DFT+U and HSE06 (see SI Figure \ref{fig:SI_HSE}), so we use DFT+U with a U of 4 eV on the Nb-\textit{d} states due to the increased computational cost of  hybrid functionals. 

 Amorphous configurations were generated from high-temperature \textit{ab initio} molecular dynamics implemented within VASP. We used scikit-learn to perform the statistical analysis with a Random Forest Classifier\cite{scikit-learn}, cross-validation was performed with the \textit{KFold} module in scikit-learn with 5 folds, and features were generated with Pymatgen~\cite{pymatgen}. The formation of YSR states and the associated impedance of the superconducting state in the SRF limit was calculated using our own code which is available on github\footnote{\url{https://github.com/Evan1415/impedance}}.

\section{Results}

\subsection{First-principles calculations of Nb oxides}

\subsubsection{Crystalline Phases: Nb$_2$O$_5$, NbO$_2$, and NbO}

We first performed DFT calculations of the electronic and magnetic properties of crystalline phases; NbO, the two phases of NbO$_{2}$ and three polymorphs of Nb$_{2}$O$_{5}$. As summarized in Figure~\ref{fig:crystalline_fig_1}, NbO and NbO$_{2}$ have partially filled \textit{d} manifolds resulting in paramagnetism. Magnetic susceptibility measurements of NbO show it to be a weakly paramagnetic metal~\cite{pollard_thesis}, consistent with this nominal $d^{3}$ oxidation state. Our DFT calculations suggest a paramagnetic metal ground state with the spin-polarized calculation being 1 meV/formula unit (f.u.) lower in energy than the non-spin-polarized case. We calculate a ferrimagnetic ordering to be slightly lower in energy  (0.07 meV/f.u) than a fully ferromagnetic state with locally projected magnetic moments of 0.06, -0.05, 0.02  $\mu_{B}$ on the three Nb in the unit cell. Finally, we investigate the influence of hydrostatic pressure on the magnetic properties by uniformly scaling the lattice parameters and allowing the internal atomic positions to optimize any residual forces. We find that negative pressure (a proxy for tensile strain) increases the tendency to spin polarizee in NbO -- the ferrimagnetic calculation becomes 3.3 meV/f.u. lower in energy than a non-magnetic case for 4\% negative pressure. We also find the largest local spin moment increases from 0.06  $\mu_{B}$  at equilibrium pressure to 0.14  $\mu_{B}$ at 4\% negative pressure as a result of the increased localization of the Nb-\textit{d} orbitals from the extended Nb-O bonds. 

NbO$_{2}$ exists in two stable structures: a high-temperature metallic rutile phase that undergoes a metal-to-insulator transition to a tetragonal insulator, which has a dilute density of paramagnetic moments~\cite{nbo2_mag_susc}. In the rutile phase, our DFT calculation converges to a non-spin-polarized ground state at equilibrium volume. However, this is likely due to the high symmetry imposed in the primitive cell calculation which does not allow for highly-disordered site-specific magnetic moments consistent with paramagnetism\cite{trimarchi_prb_18}. On the other hand, we calculate the insulating tetragonal phase to have an antiferromagnetically ordered ground state at 0 K, which is 
155 meV/f.u. lower in energy than the non-spin-polarizeed case, with a local magnetic moment of 0.93 $\mu_{B}$ at equilibrium. Similar to NbO, we find that hydrostatic pressure causes this moment to increase slightly from 0.93 $\mu_{B}$ at equilibrium pressure to 0.96 $\mu_{B}$ at 4\% negative pressure.

\begin{figure}
    \centering
    \includegraphics{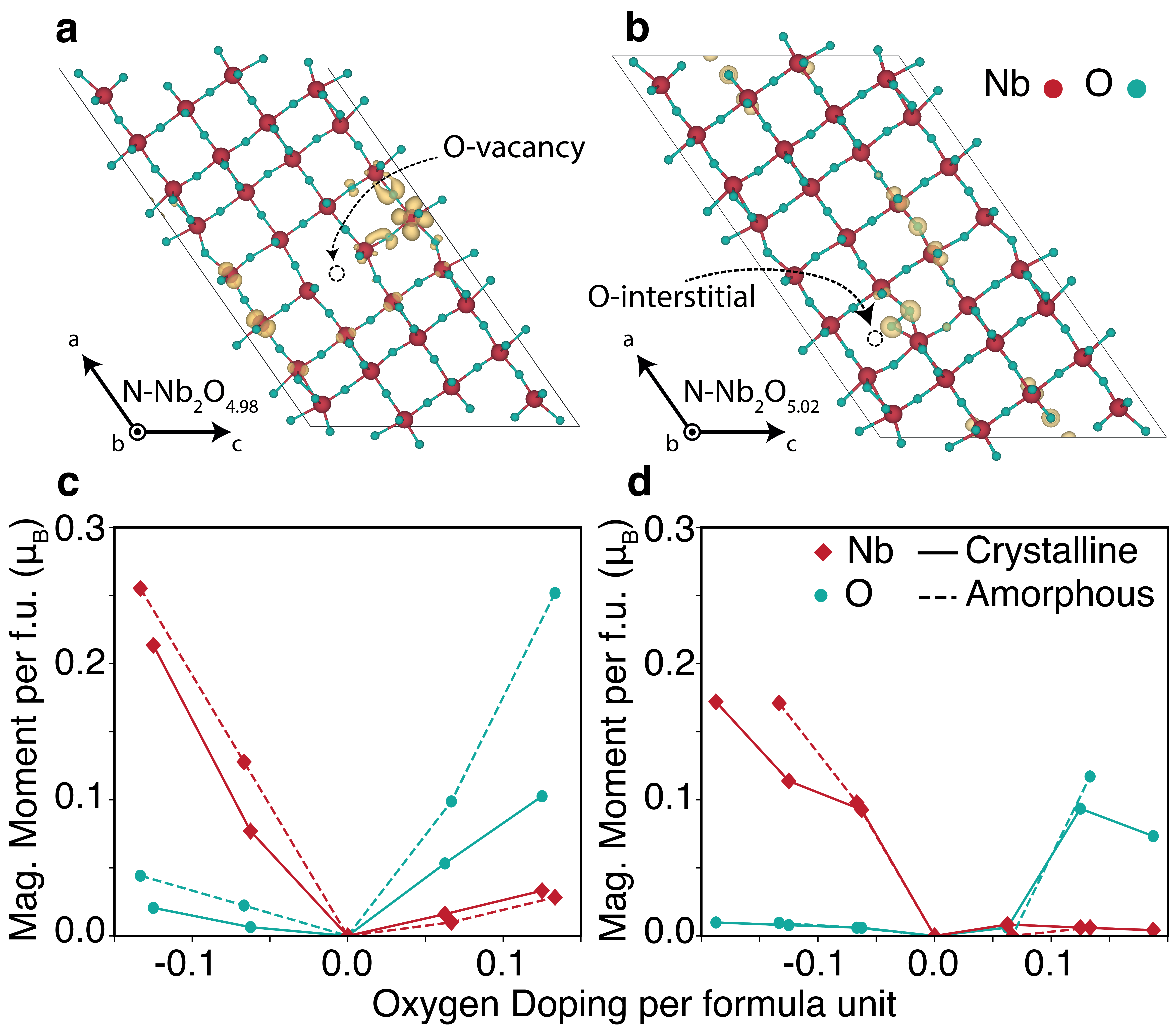}
    \caption{(a) Real space charge density projection of mid-gap states for N-Nb$_2$O$_5$ in the case of (a) an O-vacancy and (b) an O-interstitial. The calculated magnetic moments per formula unit (per Nb$_2$O$_5$) are shown as a function of the change in O stoichiometry (defined as the value of $x$ in Nb$_2$O$_{5+x}$) for the rigid band approximation (c) and explicit O doping (d) for the crystalline N-Nb$_2$O$_5$ and the amorphous Nb$_2$O$_5$ (configuration~5) structure.}
    \label{fig:amorph_dope}
\end{figure}

To explore the role of off-stoichiometry on the electronic and magnetic properties of the crystalline structures, we use both a rigid-band approximation and explicit doping cases. Firstly, we include electron and hole doping using a rigid-band model which add a compensating background charge to the calculation with our results for the N-Nb$_{2}$O$_{5}$ plotted in Figure~\ref{fig:amorph_dope}. For the stoichiometric case with Nb$^{5+}$ having an empty \textit{d}-manifold, and fully filled O-\textit{p} we find no unpaired spins as expected. However, electron doping the system shifts the Fermi level into the bottom of the conduction band, which comprises Nb-\textit{d} states, resulting in non-zero magnetic moments on Nb. Likewise, for hole doping, we find a moment is  induced on the O-\textit{p} states, consistent with the Fermi level shift into the valence band. We find similar qualitative trends for all Nb-oxides considered (SI Figure~\ref{fig:SI_RB}).

We calculated explicit defects considering both representative O vacancies and O interstitials in the monoclinic N-Nb$_2$O$_5$ phase.  N-Nb$_2$O$_5$ is chosen for its favourable thermodynamical stability~\cite{valencia_cms_14} and variety of point- and edge-sharing octahedral linkages~\cite{nowak_chemrev_99}, providing a good point of reference to the amorphous phase. We constructed a 1x2x1 supercell of N-Nb$_2$O$_5$ containing 32 formula units and fully relaxed all internal coordinates. For each O vacancy chosen at point-sharing NbO$_6$ sites, the nominal charge of Nb decreases to  Nb$^{+4.98}$. In the case of interstitials, additional O ions were placed at the midpoint of four point-sharing NbO$_6$ octahedras which correspond to a nominal charge reduction of Nb$^{+5.02}$ per interstitial where the separation between the defect images in the supercell is $7.58$ \angstrom. Figure~\ref{fig:amorph_dope}(a) illustrates the structural rearrangement for the case of a single oxygen vacancy, where mid-gap states are projected as a real space charge density. Single O vacancies near the boundary between point- and edge-sharing octahedra result in tetrahedrally coordinated Nb-sites causing an increase the average Nb-O bond length. A spin-polarizeed charge density (dangling bond) is then localized on the Nb-\textit{d} states near the vacancy. For a single interstitial oxygen as shown in Figure~\ref{fig:amorph_dope}(b), the charge density localizes on the dangling O 2\textit{p} state. In fact, the interstitial creates a creates a seven coordinated Nb bonded environment where the average Nb-O bond length is the same as the stoichiometric scenario. 

We next calculated the total and orbital resolved density of states (DOS) in Figure \ref{fig:crystalline_fig_2}. For the stoichiometric N-Nb$_{2}$O${_5}$, we confirm a diamagnetic electronic structure comprising a filled Nb-\textit{d} manifold with a bandgap of 2.74 eV. Including an oxygen vacancy donates two electrons which localize on the Nb atoms, creating Nb-\textit{d} impurity states within the gap (Figure~\ref{fig:crystalline_fig_2}, top panel).  These localized magnetic impurities reduce the overall bandgap between the top of \textit{d}-channel impurity states and the bottom of the conduction band Nb-\textit{d} states to 1.73 eV.  For an interstitial O, the resulting hole doping causes O-\textit{p} states to form within the gap, reducing the bandgap between O-\textit{p} localized impurity states and bulk Nb-\textit{d} states to 1 eV. Finally, we calculate that the doping-dependent exchange interaction, estimated by the difference between the magnetic and non-magnetic configurations. For Nb$_2$O$_{5.02}$  it is $J_{d} = 45.8$ meV , while for Nb$_2$O$_{4.98}$ we find that $J_p= 8.5$ meV, approximately six times smaller than the \textit{d}-channel case.

To better understand the defect-induced spin polarizeation, we calculate a net magnetic polarizeation as:
\begin{equation}
P(E) = [N_{\uparrow}(E)-N_{\downarrow}(E)]/[N_{\uparrow}(E)+N_{\downarrow}(E)],
\end{equation}
where $N(E)$ is the density of states of majority $\uparrow$ or minority $\downarrow$ electrons (Figure~\ref{fig:crystalline_fig_2}(b)). We find that O vacancies favour a strong spin polarizeation in the \textit{d}-channel, while O interstitials promote \textit{p}-channel spin polarizeed electronic states, consistent with our real space charge density projection of mid-gap states in Figure~\ref{fig:amorph_dope}. We find similar qualitative behaviour of the formation of localized \textit{d}-channel and \textit{p}-channel midgap states in other pentoxides, however for the case of defects in a 2$\times$2$\times$2 supercell of monoclinic B-Nb$_2$O$_5$ magnetism  in the \textit{p}-channel is suppressed. Such discrepancies can be attributed to the differing structural characteristics in the local bonding environments of the Nb pentoxides. Thus, the key structural motif for a robust magnetic oxide is maximally separated under-coordinated Nb sites, which also plays a key role for amorphous Nb$_{2}$O$_{5}$. 

\subsubsection{Amorphous Nb$_2$O$_5$}
We now turn to amorphous Nb$_{2}$O$_{5}$, which is suggested to be the dominant stoichiometry and structural makeup of the Nb-oxides thin films from experiment\cite{altoe_arxiv20, Structure_Paper}. We generate amorphous Nb$_{2}$O$_{5}$ structures using quenched molecular dynamics, with details given in the SI Section~\ref{sec:si_amorph}, and all subsequent discussions are for nine such generated amorphous structures. Importantly, we find that our \textit{ab initio} generated structures are consistent with the Nb-Nb distances observed in the fluctuation electron microscopy measurements, and consistent with X-ray absorption spectra\cite{Structure_Paper}. 

In Figure~\ref{fig:crystalline_fig_1} (far right panel), we show an example of a quenched amorphous structure, which has a combination of 4-, 5- and 6-coordinated Nb. The mean coordination number of the amorphous Nb atoms in all stoichiometric configurations is 5.62 with a standard deviation of 0.70, while the Nb atoms in the N-, M-, and B-phases have a coordination number of 5.88, 5.50, and 6, respectively. Thus, the Nb atoms are less coordinated in the amorphous phase than the crystalline phases, with the exception of the M-phase, which has significant tetrahedral Nb coordination. The calculated average density of the amorphous phase is 4.21~g/cm$^3$, which agrees well with the experimentally determined density (4.28~g/cm$^3$)~\cite{venkataraj_jap_02}. This amorphous density is very similar to the N- and M-phases (4.29~g/cm$^3$ and 4.30~g/cm$^3$, respectively), while the B-phase is significantly denser (5.29~g/cm$^3$). We find that this trend in density correlates well with doubly coordinated oxygen, which we is 60\%, 70\%, 72.5\%, for the B-, N-, and M-phases, respectively. With the average doubly coordinated oxygen  being 66.7\% for the amorphous phase, we study the N-phase in more detail given its similarity both in density and oxygen coordination.

\begin{figure}
    \centering
    \includegraphics[width=1.0\linewidth]{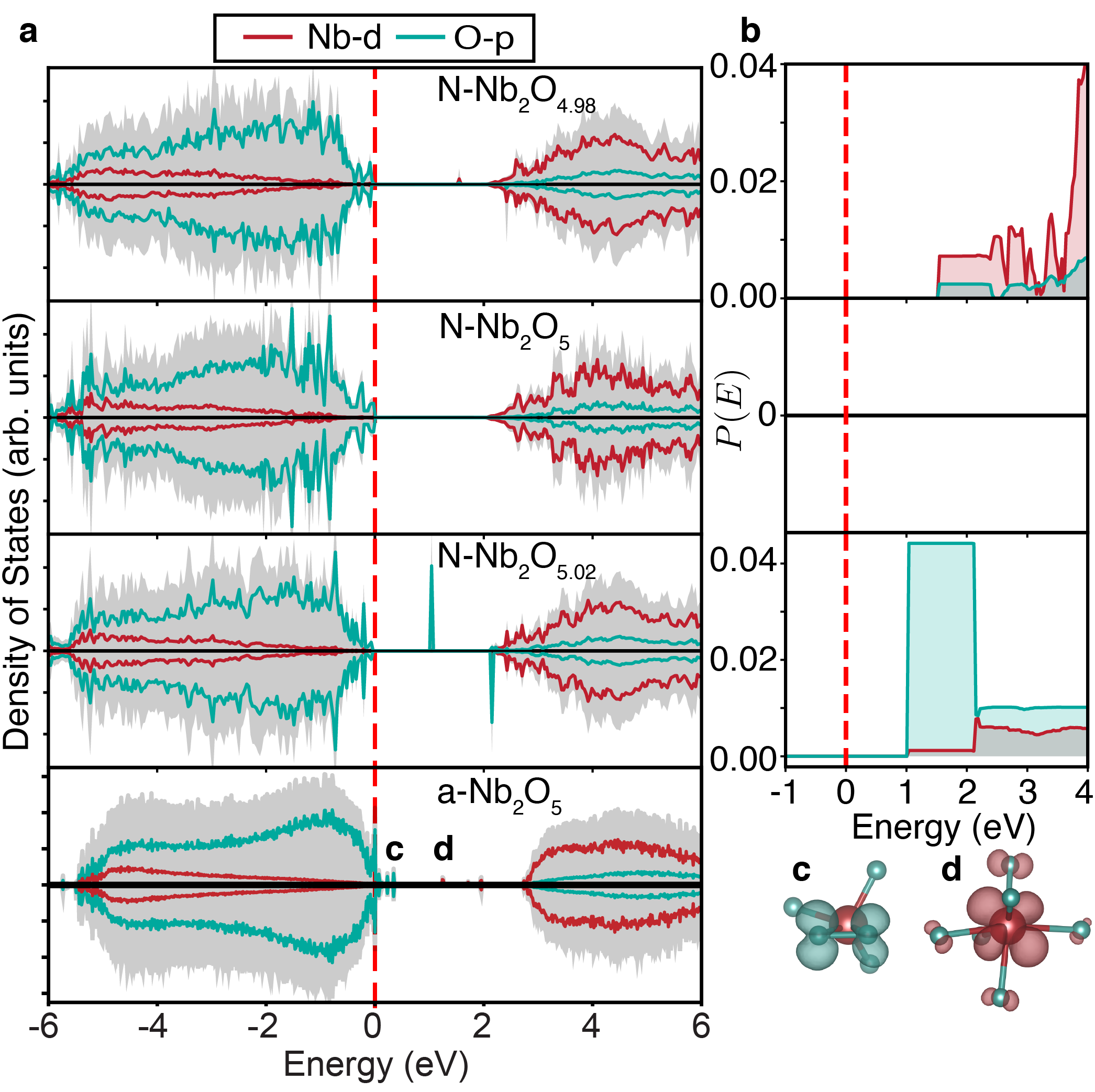}
    \caption{(a) Spin-polarizeed density of states for crystalline N-Nb$_2$O$_5$ for dilute electron-doped (N-Nb$_{2}$O$_{4.98}$), stoichiometric (N-Nb$_{2}$O$_{5}$), and hole-doped (Nb$_{2}$O$_{5.02}$) crystalline, and averaged amorphous Nb$_2$O$_5$ structures. The total DOS (shaded) is plotted along with the orbital projected states of Nb-\textit{d} states (red) and O-\textit{p} states (cyan). The Fermi level is marked with a dashed line in all cases and set to 0 eV. (b) The corresponding atomic resolved magnetic polarizeation $P(E)$ (see text for definition) of N-Nb$_{2}$O$_{4.98}$ and N-Nb$_{2}$O$_{5.02}$. (c) and (d) Charge density isosurfaces are shown for midgap O-\textit{p} and Nb-\textit{d} orbitals from amorphous calculations.}
    \label{fig:crystalline_fig_2}
\end{figure}

We calculated the spin-polarizeed DOS for amorphous Nb$_{2}$O$_{5}$ by averaging over the DOS of the nine configurations with the result plotted in Figure~\ref{fig:crystalline_fig_1}(a). To account for vacuum level shifts between  different configurations, we matched the valence band edges for each of the nine structures (ignoring midgap states). We find that our calculated bandgaps of Nb$_2$O$_5$ phases (amorphous is 2.72~eV, N is 2.10~eV, M is 2.15~eV, and B is 2.89~eV) consistent with the ranges reported in experiment given the expected underestimation of PBE bandgaps\cite{Sathasivam_et_al}. In the amorphous phase we define the bandgap as the energy gap between the top of the bulk valence band and the bottom of the bulk conduction band, neglecting midgap states. We find three classes of midgap states in our amorphous configurations of Nb$_2$O$_5$: non-magnetic O \textit{p}-states, magnetic Nb \textit{d}-states, and non-magnetic Nb \textit{d}-states. We plot the real-space projection of the example midgap states in Figure~\ref{fig:crystalline_fig_2} representing (c) non-magnetic \textit{p}-orbitals, and (d) magnetic \textit{d}-orbitals. The midgap O \textit{p}-states comprise of bonded adjacent O atoms in the structure with zero magnetic moment. The magnetic Nb \textit{d}-state in Figure~\ref{fig:crystalline_fig_1}(d) has a projected moment of $1.65$~$\mu_B$ that is highly localized on the Nb atom, similar to what is found with Nb dangling bonds in the crystalline case previously discussed. In fact, for the Nb sites where local spin moments form, we find the average coordination of Nb to be 5.5, which is lower than the average coordination number of all Nb ions in the dataset (5.62). The average Nb-O bond length for these five Nb atoms is 2.105~\AA, where the average Nb-O bond length for all Nb atoms in the set is 2.028~\AA\ (standard deviation of 0.0148~\AA). Therefore in the amorphous configurations, we find local spin moments formed where the coordination number is lower, and hence the average Nb-O bond length is longer, resulting in increased Nb orbital localization.

Two of the nine amorphous structures have spin-polarizeed ground states with local unpaired spins as discussed above. However, the overall energy of the two magnetic configurations is at least 1~eV greater than the seven other configurations (see SI Section~\ref{sec:si_amorph}), indicating that the spin-polarized configurations are a result of local minima in the material's potential energy. In the case where structures with magnetic moments are kinetically trapped, we estimate the expected magnetic moment by averaging over all considered configurations, which we find to be $0.0296$~$\mu_B$ per formula unit. We note that the calculated magnetic moment per formula unit depends on the radius of the sphere over which the projected moments are integrated, which introduces a systematic uncertainty in this result. We find that the sum of the projected moments is smaller than the total observed moment in the entire primitive cell in all amorphous configurations, suggesting that the value of $0.0296$~$\mu_B$ per formula unit is an underestimation. Prior experiments indicate that the observed magnetic moment per f.u. for the T- and TT-phase of Nb$_2$O$_5$ with a dilute concentration of O vacancies is between 0.00349-0.0164~$\mu_B$ per formula unit~\cite{herval_15}. This is comparable to, but lower than, our amorphous predictions which is explained by the increased disorder in the amorphous phase. 

We now consider deviations from stoichiometry in  amorphous Nb$_2$O$_{5}$ using both the rigid band approximation and explicit doping calculations. (Full details given in the SI Section~\ref{sec:si_dft_details}). With a rigid band approximation in Figure \ref{fig:amorph_dope}(c), we find that any deviation from stoichiometry induces local moments in a previously non-magnetic configuration, as in the crystalline case. Explicitly considering O vacancies and interstitials has similar qualitative behaviour where we find that any O vacancies or two O interstitials produce magnetic impurities in a previously non-magnetic configuration. Electron doping (or explicitly including O vacancies) creates spins localized on Nb atoms, while hole doping (O interstitials) result in local spins on O atoms. In the explicit doping case this influence is less pronounced than in the rigid band approximation as structural rearrangements around the defect suppresses the creation of local spin polarizeation.  In summary, we find that deviations from stoichiometry in both crystalline and amorphous Nb$_{2}$O$_{5}$ that result in the non-zero Nb-\textit{d} manifold, induce local moments similar in magnitude.

\begin{table}[]
    \centering
    \begin{tabular}{|c|c|}
        \hline
        Local Descriptor & Feature Importance \\ \hline
        Minimum Bond Distance & 28.0\% \\
        Maximum Bond Distance & 19.5\% \\
        Average Bond Distance & 16.2\% \\
        Polyhedron Volume & 13.2\% \\
        Std. Dev. of Polyhedron Components & 11.7\% \\
        Coordination & 11.4\% \\ \hline
    \end{tabular}
    \caption{Feature importance of local descriptors used to build the random forest classification model.}
    \label{tab:rf_importances}
\end{table}

Finally, we investigate correlations between local structural descriptors (e.g. bond lengths, coordination number and polyhedral volume) and the formation of local spin moments using a random forest classifier, with details given in the SI Section~\ref{sec:si_ml}. Confirming our insights in assessing the coordination number and bond lengths surrounding our set of configurations with local moments above, we find that the minimum bond distance (a bond length above a threshold) is the most important descriptor for determining the presence of a magnetic moment (Table~\ref{tab:rf_importances}). Inspection of individual decision trees reveals that ions with a minimum bond length above a value set by the decision tree are categorised as magnetic, which suggests that covalent bonding suppresses the formation of magnetic moments. In addition to the local bonding descriptors, we find that polyhedron shape and coordination are important secondary indicators of magnetism.


\subsection{\textit{Ab-initio} Shiba theory for surface impedance}

To connect atomic-scale disorder and defects to resonator performance, we solve for the superconducting DOS in the presence of magnetic impurities with parameters and structural information derived from our \textit{ab initio} calculations. From this updated superconducting DOS, we calculate the resulting response function to estimate the impedance changes as a result of our specific profile of magnetic impurities.

We first develop an \textit{ab initio} Shiba theory to describe the interaction of dilute spins in a superconducting host. In our Shiba model, we treat magnetic impurities as classical spins that are randomly distributed, unpolarizeed, and of finite concentration in a superconducting host. The impurity problem is solved within the infinite dimensional limit and for a finite exchange coupling $J$, where the self energy of the lattice becomes momentum independent allowing for a full summation of the perturbation series. The quasi-classical BCS Green's function for the impurity problem is
\begin{equation}
\mathcal{G}(v(\varepsilon))=\frac{v(\varepsilon)^2}{\sqrt{1-v(\varepsilon)^2}},
\label{eq:qc_gf}
\end{equation}
where $v(\varepsilon)$ is a self-consistent parameter obtained through solving 
\begin{equation}
\phi(v)=\frac{\varepsilon}{|\Delta|} = v\left(1 - \frac{1}{\tau_{s} |\Delta|}  \frac{\sqrt{1-v^2}}{((1-\gamma)^2-v^)} \right),
\label{eq:shiba}
\end{equation}
and $|\Delta|$ is the superconducting gap order parameter. $\tau_s$ is the scattering time on magnetic impurities, $\varepsilon$ is the energy and $\gamma \propto (1-J^2)/(1+J^2) + 1$ is the parameter of the Shiba theory that quantifies the magnetic exchange coupling. Parametrizing Equation~(\ref{eq:shiba}) from our \textit{ab initio} calculations allows for a self-consistent solution representing the realistic microscopic energy scales participating in defect-induced decoherence. Estimates from the non-stoichiometric crystalline calculations indicate that the ratio of exchange couplings are in the range $6 \leq \gamma^{d}/\gamma^{p} \leq 9$ per u.c (unit-cell), where $\gamma^{p,d}$ is the Shiba exchange coupling constant for \textit{p}- or \textit{d}-channel magnetic impurities, calculated from $J = E^{p,d} - E_{\text{nm}}^{p,d}$, the energy difference between the magnetic and non-magnetic configurations (see SI Section~\ref{sec:si_shiba}). In addition, we calculate the ratio between the scattering time due to magnetic impurities to be $  \tau_{s}^{p} \sim  2\tau_{s}^{d} $ per u.c, finding double the number of \textit{p}-channel magnetic impurities than \textit{d}-channel on average. Using these parameter estimates from first-principles calculations, we consider a strong exchange coupling and low impurity density, i.e $\gamma=0.9$ per u.c and $\tau_s^{-1}=0.05$ per u.c for the case of \textit{d}-channel magnetic impurities. For \textit{p}-channel magnetic impurities we consider the low exchange coupling but high impurity density limit, i.e $\gamma=0.1$ per u.c and $\tau_s^{-1}=0.1$ per u.c.

\begin{figure}
    \centering
    \includegraphics[width=\linewidth]{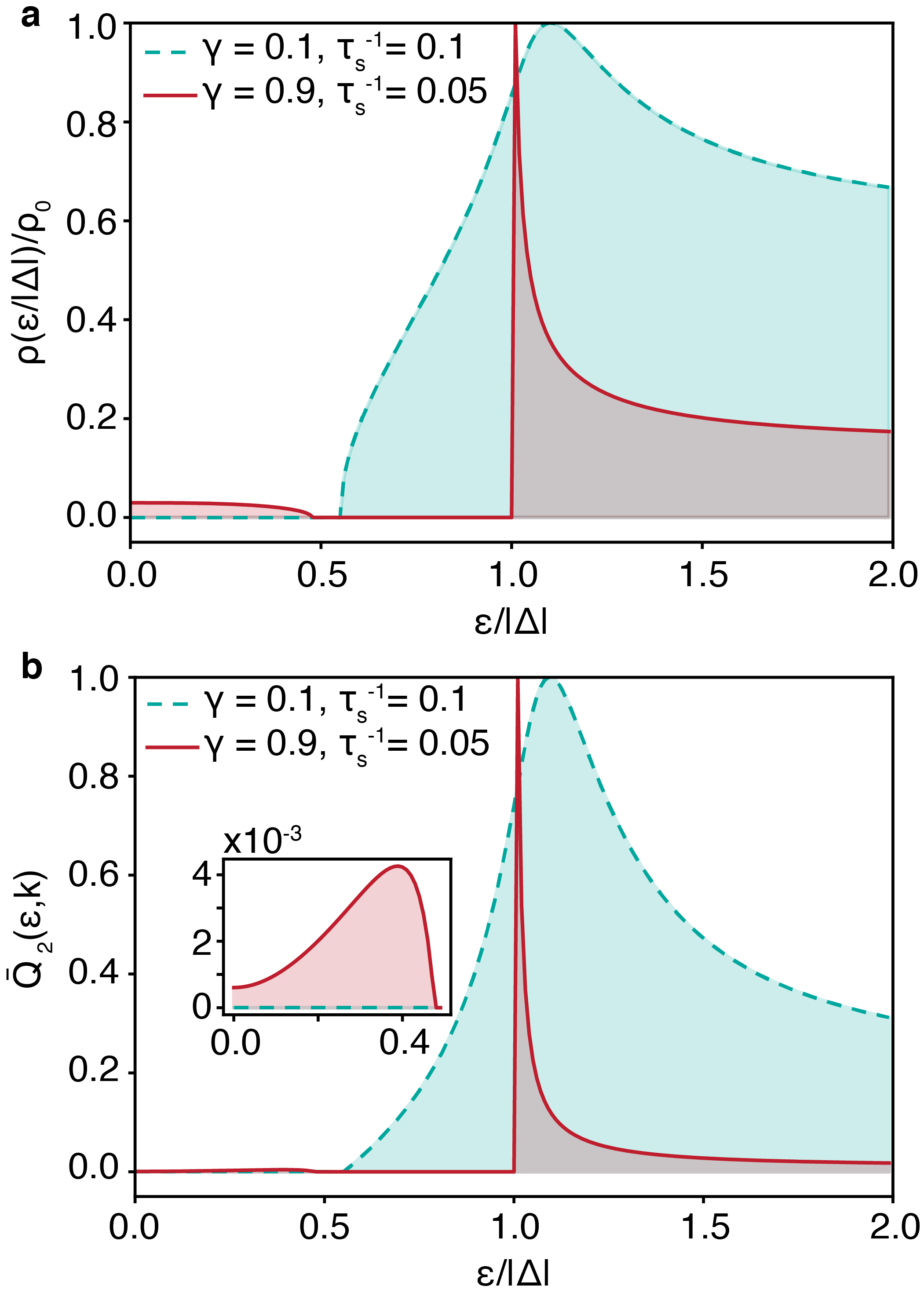}
    \caption{(a) The normalised superconducting density of states for a Nb film with magnetic impurities in the strong (red) [$\gamma=0.9,\tau_{s}^{-1}=0.05$] and weak (teal) [$\gamma=0.1,\tau_{s}^{-1}=0.1$] coupling limits in the vicinity of the Fermi level. (b) The real part of the current dissipation function in the corresponding limits, where the inset shows the region near the Fermi level which participates in the dissipation. For strong coupling, the decoherence is associated to \textit{d}-channel losses, while for weak coupling it is compatible with \textit{p}-channel losses.}
    \label{fig:shiba_dos}
\end{figure}

The resulting DOS for these limits are illustrated in  Figure~\ref{fig:shiba_dos}(a). For \textit{d}-channel magnetism an impurity band centres near the Fermi level resulting in the onset of gapless superconductivity where the impurity band is distinct from the continuum. In contrast to this, in the case of \textit{p}-channel magnetism there is no isolated impurity band as it merges with the continuum far from the Fermi level. Thus, the nature of the magnetic carriers (\textit{d}-channel or \textit{p}-channel) -- and hence the local atomic structure --  determines whether gapless superconductivity arises.

Next we explore how this distinction of \textit{p}- versus \textit{d}-channel magnetism can be useful in determining the spin contribution to decoherence in the limit of low frequency superconducting resonators. An interesting case emerges in the low frequency limit ($\omega \sim 0.01$ meV and $|\Delta| \sim 1$ meV) when the current response function can be approximated to be dependent only on its dissipative contribution \cite{pellin12}. The Fermi distribution effectively constrains only quasiparticles within $T$ of the Fermi level.  If  $\bar{Q}_{2}(\varepsilon,k) \neq 0$  then there must be a gapless superconductor, i.e  $N(\varepsilon=0) \neq 0$. Our analysis confirms this in the two limits considered (Figure~\ref{fig:shiba_dos}(b)). For \textit{d}-channel magnetic impurities, quasiparticle excitations near the Fermi level create a gapless state and cause dissipation through the finite contribution of $\bar{Q}_{2}(\varepsilon,k)$ at the Fermi level. However, for \textit{p}-channel impurities, the opposite is observed where $\bar{Q}_{2}(\varepsilon,k)$ remains zero at the Fermi level,  indicating a critical impurity density is required for \textit{p}-channel impurities to contribute to the finite dissipation at the Fermi level. In the low frequency limit the surface resistance $R(\omega)$ is quadratic in frequency\cite{pellin12}, i.e $R(\omega) \propto \omega^2$. Because of this,  gapless superconductivity -- regardless of the strength -- always results in a finite dissipation, provided there is a finite DOS at the Fermi level:
\begin{equation}
R(\omega) |_{T} = 0 \Leftrightarrow N(\varepsilon = 0) > 0. 
\end{equation}

As a result, we can attribute  an ohmic loss in superconducting films to the structural disorder and defects in the the surface oxide formed on these films.  This residual resistance is primarily associated with O deficient growth, where the primary magnetic carrier is the Nb ion. We note that while our results indicate that O surplus growth conditions do not significantly participate in the magnetic-impurity-induced decoherence processes for our Nb-oxide films, it does not rule out this is general, with details depending on the precise atomic-scale structure and chemistry of the surface layers.

In Figure~\ref{fig:SI_shiba} (SI Section~\ref{sec:si_shiba}) we present a set of plausible impurity densities that are compatible with \textit{p}-channel decoherence losses. We notice that for a considerable increase in the amount of off-stoichiometry at the surface oxide corresponding to an O surplus, then it is possible to induce \textit{p}-channel magnetic decoherence losses at an impurity density $\tau_{s}^{-1}\sim0.5$ per u.c and a higher exchange coupling of $\gamma=0.3$. As $\gamma$ is increased, so does the corresponding exchange energy between the magnetic O ions, as a result of the increased magnetic impurity density. This cooperative effect results in the onset of a gapless phase due to the broadening of the continuum into the gap, which eventually crosses the Fermi level and a concomitant dissipation. 


\section{Discussion and Conclusion}

Our DFT calculations confirm the presence of paramagnetic moments in NbO and NbO$_{2}$, as expected from their unfilled \textit{d} manifold, and consistent with experiments. However, for diamagnetic Nb$_{2}$O$_{5}$ with its $d^{0}$ configuration, we find that  structural distortions (considered here in the amorphous limit), strain, and defects can induce and enhance the formation of local spin moments. In the crystallines phases of Nb$_{2}$O$_{5}$ considered here, electron doping results in localized Nb-\textit{d} moments, while hole doping induces O-\textit{p} moments, consistent with a simple rigid-band model of shifting the chemical potential into the conduction/valence band respectively. Similar trends with electron and hole doping are also observed in the amorphous Nb$_{2}$O$_{5}$ configurations. In amorphous phases of Nb$_{2}$O$_{5}$, we find that local spin moments form even in the nomimally $d^{0}$ stoichiometry as a result of the formation of dangling bonds in the disordered phases. We observe that local spin moment formation is favored in undercoordinated Nb atoms with miminal covalent bonding, consistent with Ref.~\cite{khomskii_pnas_16} where the authors suggest that local moments are suppressed in $4d$ transition metal complexes in favor of bonding orbitals. 


We classify the local spin moments in Nb$_2$O$_5$ into two types : \textit{d}-channel states localized on Nb-\textit{d} primarily formed from O vacancies or under-coordinated Nb, and \textit{p}-channel states localized on O-\textit{p} formed from O interstitials. While the latter hole-doping scenario is difficult to achieve at equilibrium, we expect chemical and structural inhomogeneities to have local regions with excess oxygen. Furthermore, our estimates for the magnitude of the magnetic moments are comparable to other studies of wide bandgap insulating systems that exhibit defect-induced magnetism \cite{herval_15, DIM}.   

Using these atomistic calculations to inform our Shiba theory, we find that regardless of composition, \textit{d}-channel magnetic impurities have a more sizeable influence on the resulting non-TLS loss channels for Nb-resonators. This is due to the much greater \textit{d}-channel magnetic moments and exchange interaction compared to \textit{p}-channel values. While the \textit{d}-channel impurity density is on average about 50\% that of \textit{p}-channel, the exchange coupling is approximately one order of magnitude greater. Our surface impedance calculations reveal that the combination of these three factors results in a finite residual resistance for \textit{d}-channel magnetic carriers that cannot be reduced by lowering the temperature, and a dissipationless current for \textit{p}-channel impurities. Our results suggest that oxygen-deficient superconducting films, as suggested by valence band X-Ray Photoemission Spectroscopy in Ref.~\cite{Structure_Paper}, (with \textit{d}-channel magnetism) are a significant decoherence channel in Nb-based superconducting qubits. We note, however, that it is possible that \textit{p}-channel carriers also induce a dissipation, however this requires a significant increase in magnetic impurity density.


As a result, improvement of Nb-based superconducting quantum systems relies on the mitigation of the decoherence processes caused by the NbO$_x$ layer. Complete removal of the oxide layer is preferable, and may be possible through surface passivation after complete layer removal, or by changing fabrication steps to avoid the formation of the oxide layer altogether. Such surface removal and treatments were recently carried out in Nb thin films, leading to a significant improvement in quality factor of over 5 million\cite{altoe_arxiv20} and consistent with this theoretical study. Without surface oxide removal, our results suggest that annealing oxide layers to increase its crystallinity and remove oxygen vacancies will reduce the number of magnetic impurities, which agrees with the work by Proslier et. al~\cite{proslier_ieee_11}. Our work uncovers the microscopic origins of non-TLS magnetic losses in native oxides on superconducting resonators and how they influence performance. We further suggest routes to reducing the most destructive atomistic decoherence channels through fabrication procedures.

\section{Acknowledgments}
We thank John Clarke, John Vinson, Aritoki Suzuki, Adam Schwartzberg and Tess Smidt for insightful discussions and manuscript feedback. This work was funded by the U.S. Department of Energy, Office of Science, Office of Basic Energy Sciences, Materials Sciences and Engineering Division under Contract No. DE-AC02-05-CH11231 ``High-Coherence Multilayer Superconducting Structures for Large Scale Qubit Integration and Photonic Transduction program (QIS-LBNL)". 
This research used resources of the National Energy Research Scientific Computing Center (NERSC), a U.S. Department of Energy Office of Science User Facility located at Lawrence Berkeley National Laboratory, operated under Contract No. DE-AC02-05CH11231. E.S. acknowledges support from the US-Irish Fulbright Commission, the Air Force Office of Scientific Research under award number FA9550-18-1-0480 and the EPSRC Centre for Doctoral Training in Cross-Disciplinary Approaches to Non-Equilibrium Systems (EP/L015854/1). This work also used the Extreme Science and Engineering Discovery Environment (XSEDE), which is supported by National Science Foundation grant number ACI-1548562. Work at the Molecular Foundry was supported by the Office of Science, Office of Basic Energy Sciences, of the U.S. Department of Energy under Contract No. DE-AC02-05CH11231.
\bibliography{refs}

\end{document}


\title{Microscopic Theory of  Magnetic Disorder-Induced Decoherence in Superconducting Films}

\author{Evan Sheridan}
\thanks{These authors contributed equally}
\affiliation{Materials Science Division, Lawrence Berkeley National Laboratory, Berkeley, CA 94720, USA}
\affiliation{Molecular Foundry, Lawrence Berkeley National Laboratory, Berkeley, CA 94720, USA}
\affiliation{King's College London, Theory and Simulation of Condensed Matter, WC2R 2LS London, UK}

\author{Thomas F. Harrelson}
\thanks{These authors contributed equally}
\affiliation{Materials Science Division, Lawrence Berkeley National Laboratory, Berkeley, CA 94720, USA}
\affiliation{Molecular Foundry, Lawrence Berkeley National Laboratory, Berkeley, CA 94720, USA}

\author{Eric Sivonxay}
\affiliation{Energy Technologies Area, Lawrence Berkeley National Laboratory, Berkeley, CA 94720, USA}
\affiliation{Department of Materials Science and Engineerng, University of California, Berkeley, Berkeley, CA 94704, USA}

\author{Kristin A. Persson}
\affiliation{Energy Technologies Area, Lawrence Berkeley National Laboratory, Berkeley, CA 94720, USA}
\affiliation{Department of Materials Science and Engineerng, University of California, Berkeley, Berkeley, CA 94704, USA}

\author{M. Virginia P. Alto\'{e}}
\affiliation{Molecular Foundry Division, Lawrence Berkeley National Laboratory, Berkeley, CA 94720, USA}

\author{Irfan Siddiqi}
\affiliation{Department of Physics, University of California, Berkeley, California 94720, USA}
\affiliation{Center for Quantum Coherent Science, University of California, Berkeley, California 94720, USA}

\author{D. Frank Ogletree}
\affiliation{Molecular Foundry Division, Lawrence Berkeley National Laboratory, Berkeley, CA 94720, USA}

\author{David I. Santiago}
\affiliation{Center for Quantum Coherent Science, University of California, Berkeley, California 94720, USA}

\author{Sin\'{e}ad M. Griffin}
\affiliation{Materials Science Division, Lawrence Berkeley National Laboratory, Berkeley, CA 94720, USA}
\affiliation{Molecular Foundry, Lawrence Berkeley National Laboratory, Berkeley, CA 94720, USA}



\maketitle

\section{Density Functional Theory Calculation Details} \label{sec:si_dft_details}

\subsection{Convergence Parameters and Calculation Details}

\begin{figure}[h!]
    \centering
    \includegraphics[width=\linewidth]{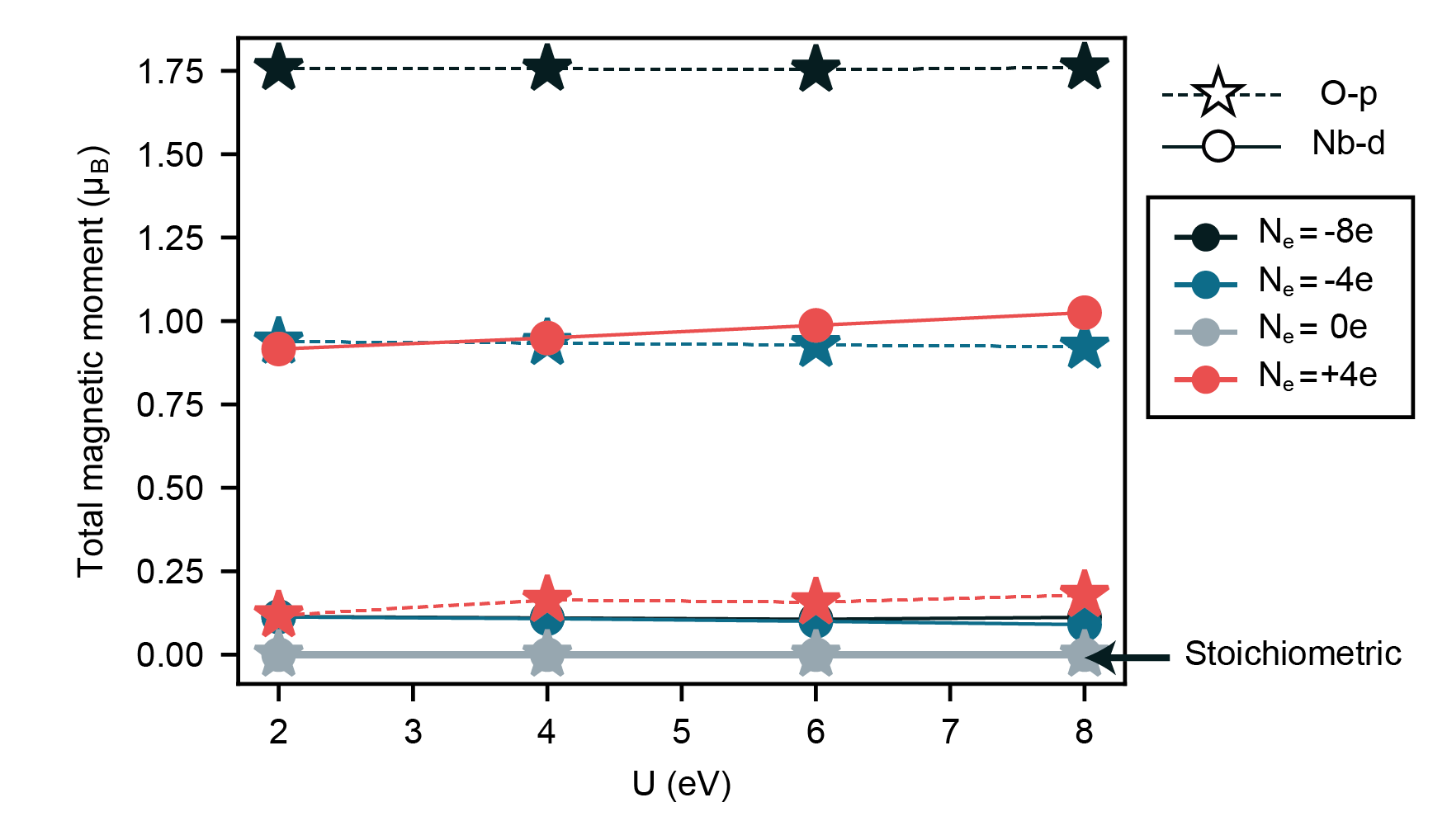}
    \caption{The dependence of the species-dependent total magnetic moment as a function of the effective Hubbard U parameter for both stoichiometric and off-stoichiometric configurations of B-\n. Electron and hole doping is included by adding a compensating background charge to the \emph{ab initio} calculation. }
    \label{fig:u_study}
\end{figure}

\begin{figure}[h!]
    \centering
    \includegraphics[width=\linewidth]{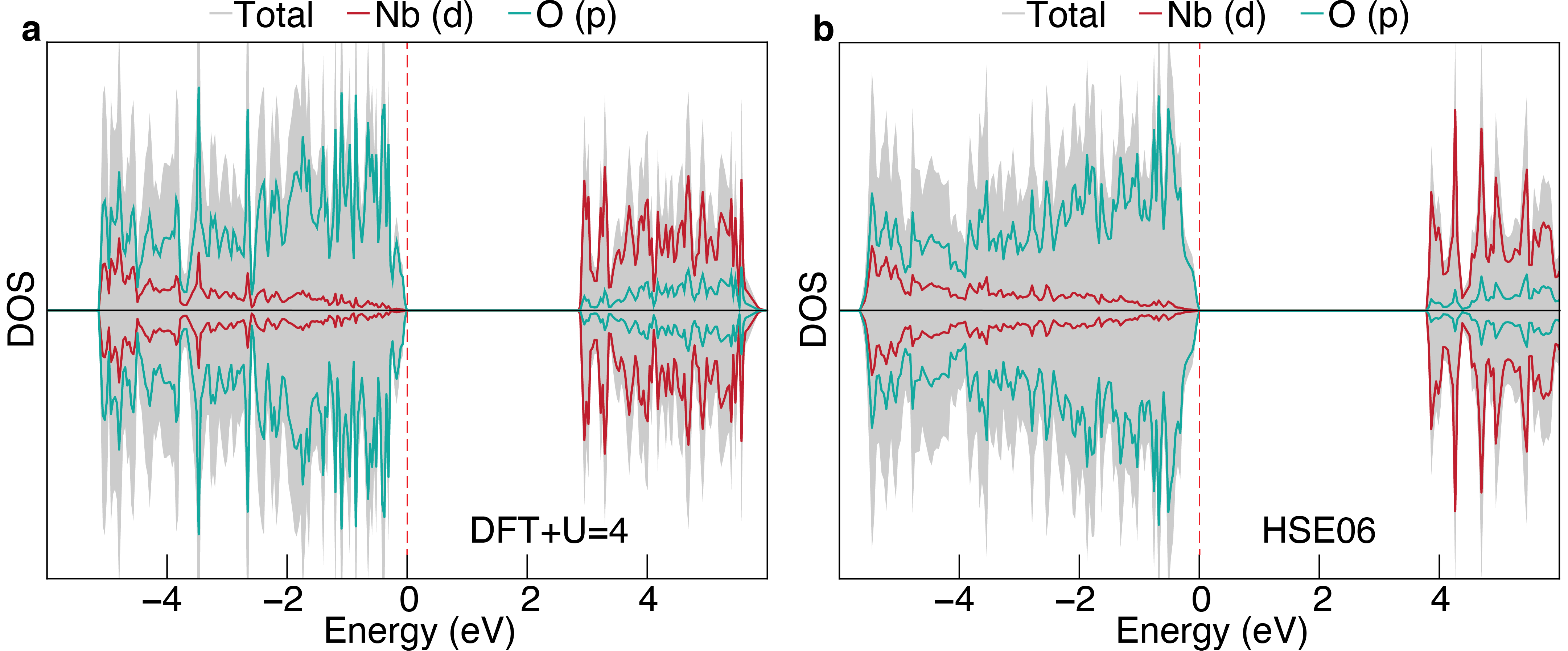}
    \caption{ Comparison between PBE+U=4 eV and HSE06 for B-Nb$_2$O$_5$. The experimental bandgap is estimated to be $3.8 - 3.9$ eV \cite{nico_pms_16}}
    \label{fig:SI_HSE}
\end{figure}

Our Density Functional Theory (DFT) calculations were performed using the Vienna Ab initio Simulation Package (VASP)~\cite{kresse93}. For all calculations, we chose the PBE functional~\cite{PBE} with a U correction of 4~eV on the Nb \textit{d}-orbitals, consistent with other DFT studies of Nb oxides~\cite{kocer_prb19}, and justified in the main text. The sensitivity of our calculated magnetic moments as a function of the effective Hubbard U parameter was investigated (Figure~\ref{fig:u_study}) for all stoichiometries considered. Additionally, the choice of U value is validated against the HSE06 functional in Figure~\ref{fig:SI_HSE}. The plane-wave cutoff was 600~eV for all calculations, and we used the projector augmented wave (PAW) pseudopotentials~\cite{kresse99} including Nb 4\textit{pd} 5\textit{s}, O 2\textit{s} 2\textit{p} electrons as valence. 


\begin{figure*}
    \centering 
    \includegraphics[width=\linewidth]{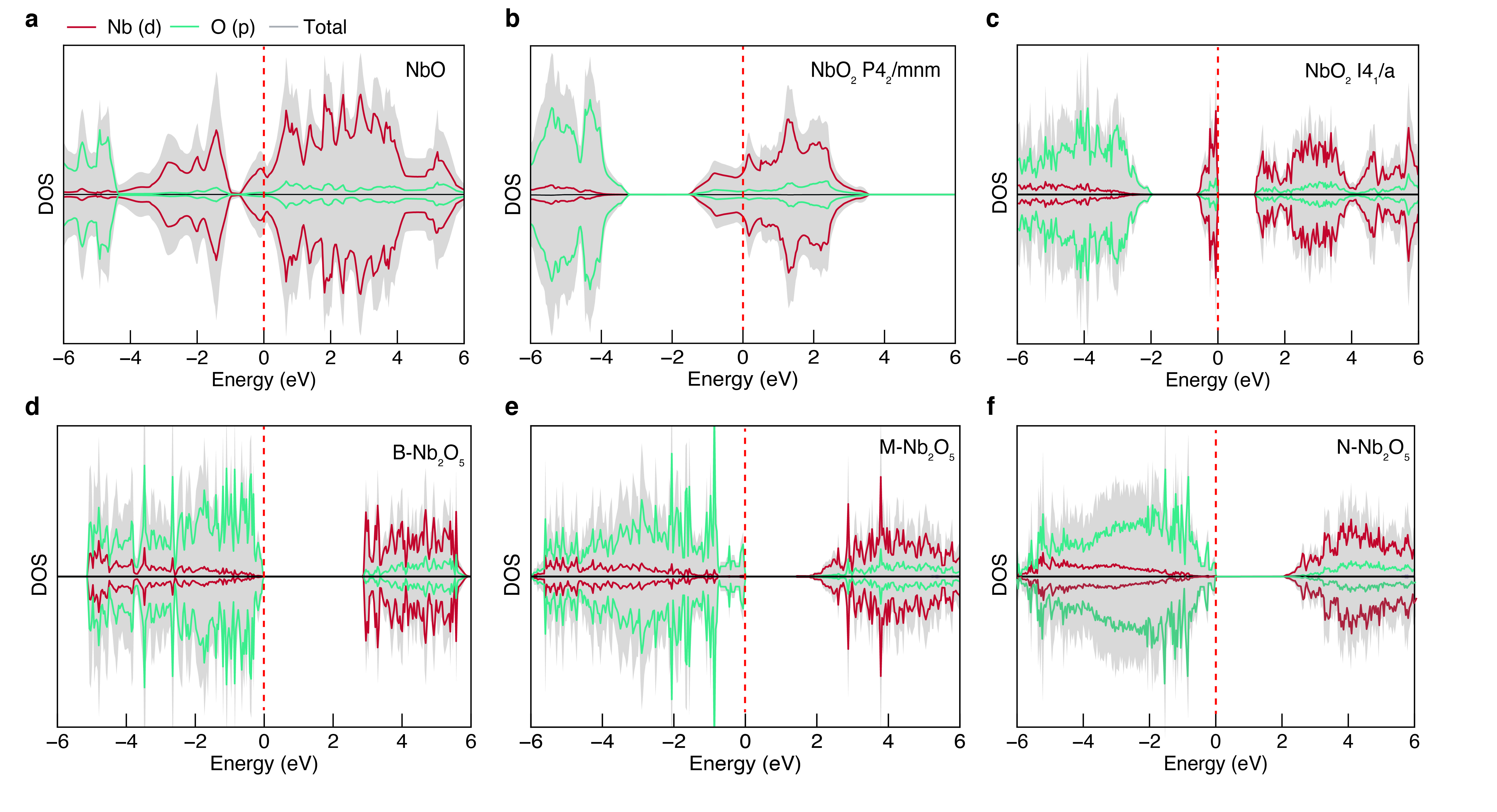}
    \caption{The total and orbital-projected density of states for the (a) NbO (\emph{Pm3m)},  (b) NbO$_2$ (\emph{P4$_2$/mnm}), (c) NbO$_2$ (\emph{I4$_1$/a}), (d) B-Nb$_2$O$_5$, (e) M-Nb$_2$O$_5$ and (f) N-Nb$_2$O$_5$. The Fermi level is indicated by the dashed red line. Nb (d) states are illustrated by the solid red line, O (p) by the solid teal line and the Total Density of states by the grey shaded area.} 
    \label{fig:SI_suboxide_dos}
\end{figure*}

For the crystalline calculations, the k-point grid was chosen such that the self-consistent field energy was converged to 10$^{-7}$~eV and forces were converged to 1~meV\AA$^{-1}$ (Table~\ref{tab:crys_VASP}). The initial structures for crystalline NbO, NbO$_2$, and Nb$_2$O$_5$ were obtained from experiments\cite{nico_jpcc_11, bowman_66, nico_pms_16, nowak_chemrev_99}, and the resulting calculated structural and electronic properties are summarized in Table~\ref{tab:crys_details} and in Figure~\ref{fig:SI_suboxide_dos}. 

Amorphous structures were generated with \textit{ab initio} molecular dynamics within VASP. A cell of 30 Nb atoms and 75 O atoms was melted at 5000~K for 20~ps with the PBE functional using the NpT ensemble. From these we selected configurations that were separated by 2~ps, and we optimized the atomic positions, lattice constants, and angles for each amorphous configuration using a force cutoff of 0.01~eV\AA$^{-1}$. The initial geometry optimisation was performed without a U value  on the Nb-\textit{d} states. Following these initial geometry optimisations, we re-optimised the structures with a U value of 4~eV on Nb-\textit{d} orbitals using the same force convergence cutoff. For each calculation, we performed spin-polarized calculations with the Nb moments initialised to 4 $\mu_B$. 

\begin{table}[h!]
    \centering
    \begin{tabular}{| c | c |}
      \hline
      Material & K-point grid \\ \hline
      NbO (\emph{Pm3m}) & $6 \times 6 \times 6$ \\
      NbO$_2$ (\emph{P4$_2$/mnm}) & $6 \times 6 \times 6$ \\
      NbO$_2$ (\emph{I4$_1$/a}) & $3 \times 3 \times 6$ \\
      Nb$_2$O$_5$ (M) & $2 \times 2 \times 12$ \\
      Nb$_2$O$_5$ (N) &  $2 \times 7 \times 2 $ \\
      Nb$_2$O$_5$ (B) &  $2 \times 10 \times 8$ \\
      Nb$_2$O$_5$ (amorphous) & $3 \times 3 \times 3$ \\
      \hline
    \end{tabular}
    \caption{Brillouin zone sampling (k-point) grids used for the crystalline Nb-oxide calculations.}
    \label{tab:crys_VASP}
\end{table}


\begin{table*}\centering
  \ra{0.2}
  \begin{tabular}{|c|c|c|c|c|c|c|c|}
    \hline
    \rowcolor{black!20}
    Form & Phase & Spacegroup & DFT Lattice parameters  & Expt. Lattice parameters & W (eV) &  $E_{g}$\textsuperscript{expt.} (eV) & $E_{g}$\textsuperscript{theory} (eV) \\ \hline
    \no & cubic & \emph{Pm3m} & \makecell{$a = 4.26$ \angstrom}   & \makecell{$a = 4.21$  \angstrom}&  10 & 0.0  & 0.0 \\ \hline
    \nt & tetragonal & \emph{P4$_2$/mnm} & \makecell{$a = 5.0$  \angstrom  \\ $c = 2.93$  \angstrom }  & \makecell{$a = 4.50$   \angstrom  \\ $c = 2.86$  \angstrom }& 6.0 & 0  & 0 \\ \hline
    \nt & tetragonal & \emph{I4$_1$/a} & \makecell{$a = 13.89 $  \angstrom \\ $c = 6.10$  \angstrom}  & \makecell{$a = 13.70$ \angstrom  \\ $c = 5.98$  \angstrom }& 6.0 &  0.5 & 1.1 \\ \hline
    B-\n & monoclinic & \emph{C2/c} & \makecell{$a = 13.0$  \angstrom  \\ $b = 4.94$  \angstrom  $ \beta = 103.54\degree $\\ $c = 5.61$  \angstrom} & \makecell{$a = 12.73$  \angstrom  \\ $b = 4.88$  \angstrom  $ \beta = 105.1\degree$ \\ $c = 5.56$  \angstrom } &3.15 & 3.79  & 2.89 \\ \hline
    N-\n & monoclinic & \emph{C2/m} & \makecell{$a = 29.38$  \angstrom  \\ $b = 3.79$  \angstrom  $ \beta = 125.1\degree $\\ $c = 18.06$  \angstrom} & \makecell{$a = 28.51$  \angstrom  \\ $b = 3.83$  \angstrom  $ \beta = 120.8\degree $ \\ $c = 17.48$  \angstrom } & 4.92 & -\footnote{The bandgap for N-\n has not been measured, as far as we are aware}  & 2.1 \\ \hline
    M-\n & tetragonal & \emph{I4/mmm} & \makecell{$a = 20.77 $  \angstrom \\ $c = 3.86$  \angstrom} & \makecell{$a = 20.44 $  \angstrom   \\ $c = 3.83$  \angstrom } &4.89 &  3.8-3.9 & 2.15 \\ \hline
    a-\n & amorphous & - & \makecell{$a = 11.69$  \angstrom \\ $b = 11.79$  \angstrom \\ $c = 11.48$  \angstrom } & \makecell{$a = 4.26$ } &6.43 & 3.4  & 2.72 \\ \hline
  \end{tabular} \label{tab:Nb2O5_polymorphs}
  \caption{Structural optimisations with DFT+U were performed with a U$_{\text{eff}}$ value of 4.0~eV on Nb \textit{d}-orbitals. W~(eV) is the bandwidth of the conduction bands and $\Delta$ denotes the band gap (eV). The experimental bandgaps $\Delta$\textsuperscript{expt.} and lattice parameters are referenced from literature~\cite{nico_pms_16}}
\label{tab:crys_details}    
\end{table*}


\subsection{Doping Calculation Details}
To assess the influence of oxygen defects on niobium suboxides we perform a series of \textit{ab initio} simulations to include charge doping using two different approaches. We first use the `rigid-band method', includes a static correction to the exchange correlation functional via a compensated background neutralizing charge calculated from the homogenous LDA electron gas, at no additional computational cost. However, for the second approach, we explicitly include defects into the structure, which gives a more accurate description of the local environment around real defects, but at a greater computational cost. 

\begin{figure}
    \centering
    \includegraphics[width=0.7\linewidth]{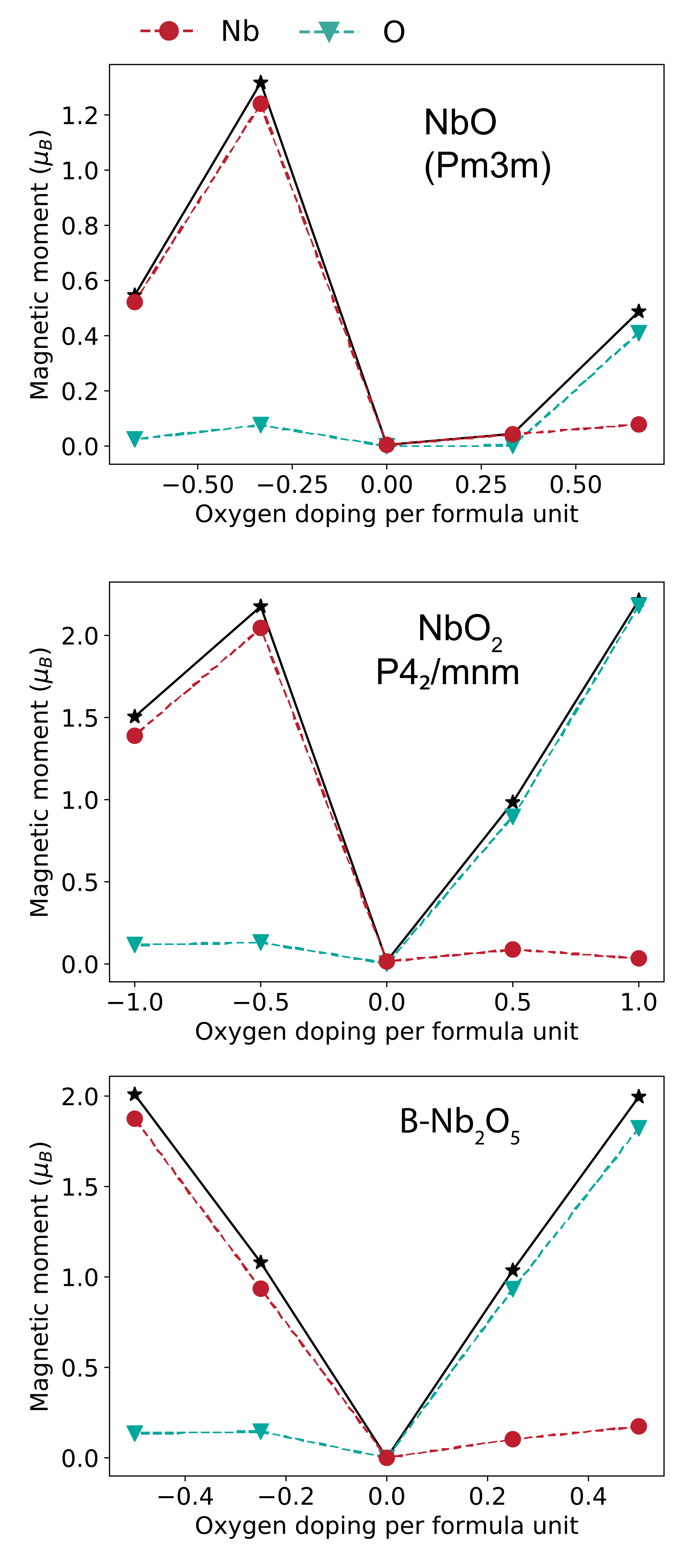}
    \caption{Rigid band doping calculations for a sample of the Nb suboxides studied, with the total magnetic moment of the unit cell on the y-axis. Oxygen doping per formula unit corresponds to the quantity of off-stoichiometric oxygen needed to remove {-8e, -4e, 0e, +4e, +8e} electrons as achieved using the rigid band model.} 
    \label{fig:SI_RB}
\end{figure}

We select a representative collection of crystalline Nb-suboxides for our rigid band calculations: NbO (\emph{Pm3m}), NbO$_2$ (\emph{P4$_2$/mnm}), B-Nb$_2$O$_5$ (\emph{C2/c}), N-Nb$_2$O$_5$ (\emph{C2/m}), and amorphous configuration no.~5. For each suboxide studied, we vary the electron doping away from the stoichiometric chemical potential by \{-8e,-4e,+4e,+8e\} per unit cell (\{-4e, -2e, +2e, +4e\} for configuration no.~5), where e is the electron charge. The results of these calculations are summarised in Figure~\ref{fig:SI_RB} and Figure~2c (main text) for N-Nb$_2$O$_5$ and amorphous configuration no.~5. In all cases, we observe the emergence of two magnetic channels depending on the type of charge doping. For electron doping, we consistently find \textit{d}-channel magnetic order, that is, induced magnetism localized on \textit{d} orbitals. For hole doping, we instead find \textit{p}-channel magnetic order. The overall magnetic moments per atomic species suggests a much larger moment residing one the Nb-\textit{d} orbitals than the O-\textit{p} orbitals, which we attribute to the tendency for magnetic character to occur on localized orbitals. We note that the differences in the number of atoms in the unit cells for N-phase (112 atoms) and B-phase (28 atoms) of crystalline Nb$_2$O$_5$, result in slightly different doping ratios per atom for the cases we consider.

Nb$_2$O$_5$ is the most thermodynamically stable Nb-suboxide and we study the influence of explicit crystalline defects (vacancies and interstitials) in this oxide only. In the N-phase we doubled the unit cell along the b-axis to avoid defect-defect interactions (see Table~\ref{tab:crys_details}) resulting in a separation of 7 \angstrom between defects, with the defect coordinates summarized in Table~\ref{tab:crys_defects}. We next included explicit defects (vacancies and interstitials) in amorphous configuration no.~5 to compare the behaviour between crystalline and amorphous phases. When creating a second vacancy, we considered two limits: removing an O nearest to the original defect, and removing an O that is well separated ($\sim6$~\angstrom) from the original defect, resulting in two new amorphous structures, each with two O vacancies. The results are summarised in Figure~2d (main text) which are compared to  our  rigid-band calculations, displayed in Figure~2c (main text). We expect similar trends to be observed for other explicit defect calculations in the other suboxides.

\begin{table}[h]
    \begin{tabular}{| c | c |}
      \hline
      Defect & Coordinates (fractional) \\ \hline
      1 O-corner sharing defect & (0.43,0.75,0.5) \\
      2 O-corner sharing defects  & (0.43,0.75,0.5), (0.45, 0.75, 0.36)\\
      1 O interstitial & (0.5,0.75,0.5)\\
      2 O interstitial & (0.5,0.75,0.5), (0.36,0.75,0.5) \\
      \hline
    \end{tabular}
    \caption{Crystalline defect details for 1x2x1 N-\n}
    \label{tab:crys_defects}
\end{table}

\subsection{Amorphous Calculations} \label{sec:si_amorph}

\subsubsection{Energetic Comparisons}
\begin{table}[h]
    \centering
    \begin{tabular}{|c|c|c|c|}
    \hline
        Configuration & $E_r$(eV) & Avg. Coord. & Local Moments\\ \hline
         1  & 4.991 & 5.80 & Y\\
         2  & 4.840 & 5.67 & N \\
        3  & 1.116 & 5.63 & N \\
        4  & 6.513 & 5.53 & Y \\
         5  & 2.453 & 5.80 & N \\
        6  & 2.335 & 6.0 & N \\
        7  & 0.884 & 5.7 & N \\
        8  & 2.972 & 5.77 & N \\
         9  & 0.000 & 5.50 & N \\\hline
    \end{tabular}
    \caption{ Calculated difference in relative energy ($E_r$) between a given configuration and the energy of configuration 9 (the lowest energy configuration in the dataset). Average coordination numbers for Nb atoms in each configuration.  A site is reported as having a local moment if it has projected moments above 0.01~$\mu_B$. }
    \label{tab:amorph_energies}
\end{table}
In Table~\ref{tab:amorph_energies}, we show the calculated ground-state energies for each amorphous configuration along with whether the resulting configuration hosts local moments. We consider a configuration to host local moments (i.e. to be paramagnetic) if it has projected moments above 0.01~$\mu_B$. We observe that the paramagnetic configurations (no. 1 and no. 4) have higher energies than  diamagnetic configurations (no local moments) in all cases. The diamagnetic configuration with the highest energy is configuration no.~2, which  has projected moment of 0.002~$\mu_B$, which we could classify as weakly paramagnetic. 

\subsubsection{Densities of States}

\begin{figure*} 
    \centering
    \includegraphics[width=\linewidth]{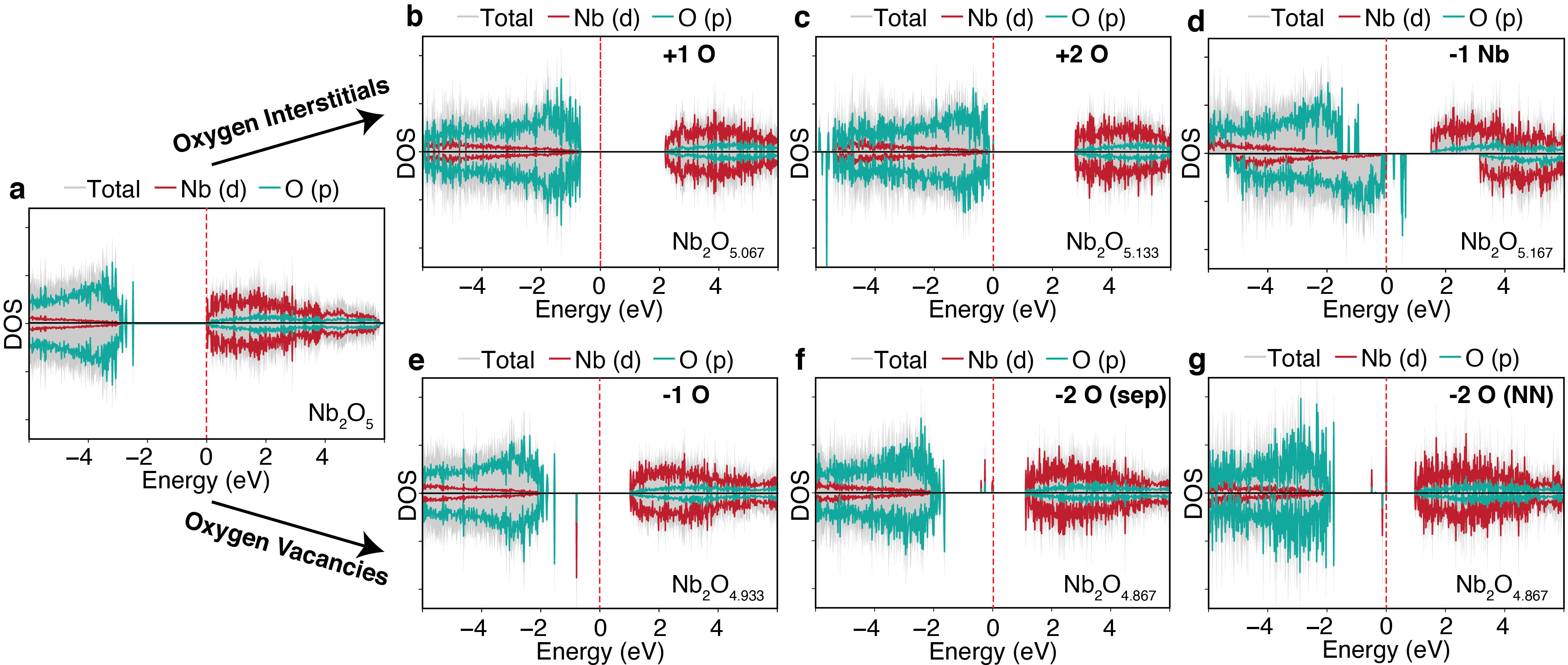}
    \caption{Calculated spin-polarized total, and projected densities of states for the amorphous configuration no.~5 structure with O and Nb defects. The total DOS is grey, the Nb projection is red, and the O projection is teal, with the Fermi level indicated by the dashed line and set to 0 eV. (a) DOS of stoichiometric Nb$_2$O$_5$. (b-d) DOS with (b) one and (c) two oxyge interstitials, and (d) a DOS with a Nb vacancy. (e-g) DOS for amorphous structure with (e) one O vacancy, and two O vacancies; one structure (f) in which the vacancies are significantly separated, and another structure (g) in which the O vacancies are nearest neighbours. Labels for the type of defect are in the upper right of each panel, and the corresponding stoichiometry is given in the bottom left of each panel.}
    \label{fig:si_amorph_dos}
\end{figure*}

\begin{figure} [h]
    \centering
    \includegraphics[width=0.95\linewidth]{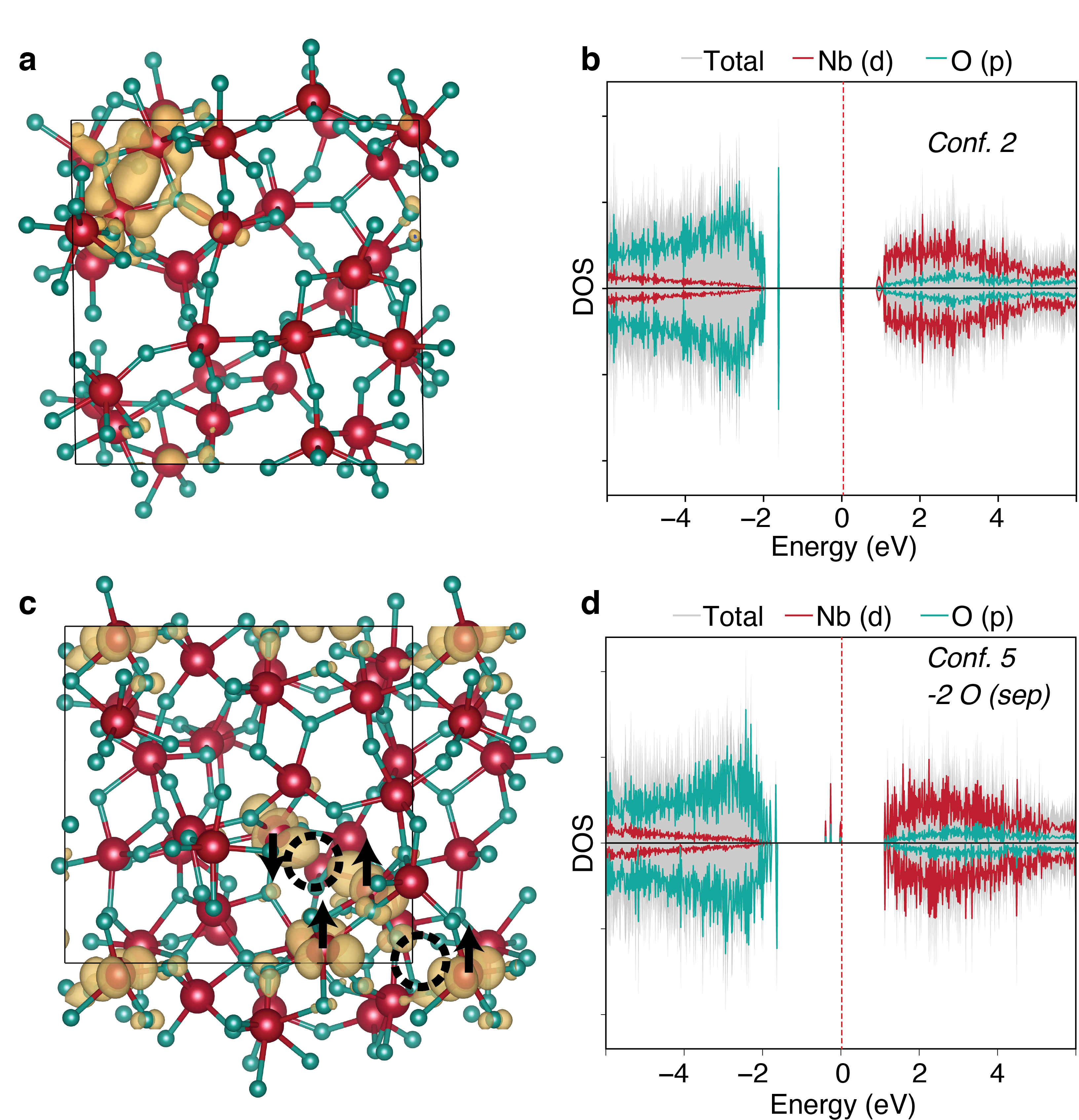}
    \caption{(a) Calculated real-space charge density of the midgap states for the two spin channels of the impurity band in configuration no.~2. (b) The density of states for configuration no.~2. (c) Calculated real-space charge density of the midgap states for the four magnetic impurity states in configuration no.~5 with two separated oxygen vacancies. The original location of the vacancies are given by dashed circles, and the projected magnetic moments are given by black arrows. (d) The density of states for configuration no.~5 with 2 separated impurities.}
    \label{fig:si_amorph_parchg}
\end{figure}

In Figure~\ref{fig:si_amorph_dos}, we plot the density of states (DOS) for each of the amorphous defect calculations. The calculations including interstitial oxygen show changes to the valence band edge, with  magnetism  contained at lower energies in \textit{p}-orbitals in the valence band. For the single O interstitial DOS (Figure~\ref{fig:si_amorph_dos}b), we see the removal of midgap \textit{p}-states, which lowers the Fermi level from the conduction band edge to the middle of the gap. For the Nb vacancy calculation, which is the same as the presence of an extra 2.5 interstitial O atoms, we observe exchange-driven spin polarization. For the O vacancies, we see the  introduction of midgap \textit{d}-states, which are magnetic. In the case of double O vacancies (Figure~\ref{fig:si_amorph_dos}f/g), we observe the presence of both spin up and spin down midgap impurity states. For the nearest neighbour vacancy case, there is only one atom with a dominant projected moment of 1.8~$\mu_B$, but we see prominent up and down spin midgap states in the DOS. Inspection of the structure in which the two O vacancies are well-separated reveals that there are four primary impurities; three up spins, and one down spin, see the partial DOS of all four impurity bands in Figure~\ref{fig:si_amorph_parchg}c. The three up spins are connected through edge sharing polyhedra, in which the Nb-O-Nb bond angle is close to 90$\degree$, indicating that the presence of superexchange. The down spin is not directly connected to the other spins through O bonds, but it is adjacent to a void that separates the down spin from an up spin, suggesting that there is some \textit{d}-\textit{d} hybridization that favours antiferromagnetic order. When the adjacent spins become too hybridized, the magnetisation becomes suppressed, as seen in the partial density of states for configuration no.~2 in Figure~\ref{fig:si_amorph_parchg}a. The DOS (Figure~\ref{fig:si_amorph_parchg}d) shows that the impurity state has spin channels at equal energies, but the projected magnetic moment on each atom is 0.002~$\mu_B$. Also, the real-space charge density of the orbital shows significant delocalization across two Nb atoms and the connecting O atoms, indicating that the two \textit{d}-states formed a bonding orbital.

\subsubsection{Energetics of Magnetic Configurations}
\begin{table}[]
    \centering
    \begin{tabular}{|c|c|c|c|c|}
    \hline
        \multirow{2}{*}{Conf.} & \multirow{2}{*}{Defect Type} & E$_0-$E$_P$ & Total Mag. Mom. & \multirow{2}{*}{Orb. type} \\ 
         & & (eV) & ($\mu_B$) & \\ \hline
        1 & None &  0.900 & 1.935 & \textit{d} \\
        4 & None & 2.25 & 2.057 & \textit{d} \\
        5 & -2 O & 0.329 & 1.685 & \textit{d} \\
        5 & -2 O (sep) & 1.873 & 3.733 & \textit{d} \\
        5 & +2 O & 0.629 & 1.849 & \textit{p} \\
        5 & -1 Nb & 1.09 & 4.785 & \textit{p} \\\hline
    \end{tabular}
    \caption{Calculated energy difference for spin polarized (E$_P$) and non-spin-polarized (E$_0$) for amorphous Nb$_2$O$_5$ structures that had converged to a spin-polarized ground state. The total spins column is the sum of the absolute value of the projected moments in the entire cell. Orbital type is the character of the magnetic impurities in the structure (either \textit{p}- or \textit{d}-type). }
    \label{tab:amorph_mag}
\end{table}

In Table~\ref{tab:amorph_mag} we summarise the results of spin-polarized and non-spin-polarized calculations of a selection of the amorphous structures. The nearest neighbour double O vacancy structure has a higher energy than the well-separated double vacancy structure by 0.93~eV, suggesting that nearest neighbour O vacancies are  energetically unfavourable. The magnetization energy for configuration no.~5 with no defects cannot be provided as a baseline, because no magnetic configuration converged. Thus, we assume that any magnetic configuration for configuration no.~5 is unfavorable. For the structures in Table~\ref{tab:amorph_mag} with approximately two spins, we estimate that the difference in energy between the calculation corresponds to the exchange coupling that stabilizes the magnetic state. For the double O vacancy calculation where the vacancies were well separated, the energy difference is the direct sum between two separate exchange interactions. Thus, we observe that the range of exchange couplings for \textit{d}-magnetic structures is between 0.9~eV and 2.25~eV. For the \textit{p}-magnetic structures, we  estimate the upper bound for the exchange coupling to be 0.629~eV. The lower bound for the exchange interaction is more difficult to estimate since the number of spin-spin interactions for the Nb vacancy is not clearly defined with 5 spins somewhat delocalized around the Nb vacancy. If we assume that the number of spin-spin interactions is 5, similar to a cyclic ring of spins around the vacancy, then the lower bound for the exchange coupling for \textit{p}-magnetic systems is 0.21~eV.

\subsubsection{Structural Analysis}
\begin{figure}
    \centering
    \includegraphics[width=0.6\linewidth]{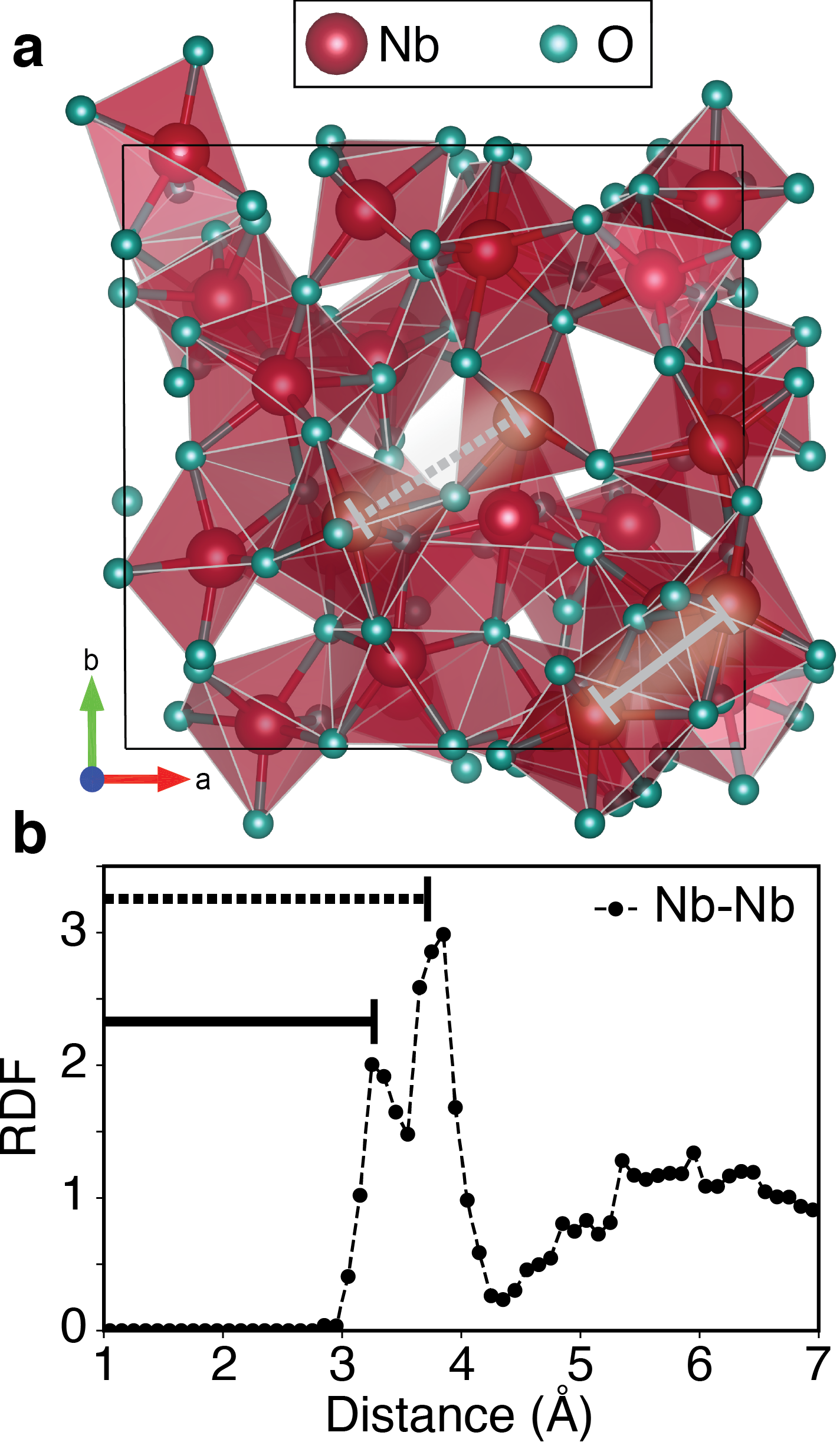}
    \caption{(a) Example of an amorphous configuration (configuration no.~5) obtained after quenching a molecular dynamics snapshot. Solid and dashed lines represent Nb-Nb distances between edge-sharing and point-sharing polyhedra, respectively. (b) Radial distribution function obtained from all quenched molecular dynamics snapshots. The peaks are labelled solid and dashed lines from panel \textit{a}.}
    \label{fig:amorph_struct}
\end{figure}

We calculated the Nb-Nb radial distribution function (RDF) averaged over all stoichiometric amorphous Nb$_2$O$_5$ configurations to understand the local bonding environment in the amorphous structures. We observe a bimodal distribution at 3.35~\AA, and 3.85~\AA, which we attribute to the presence of edge-sharing and point-sharing polyhedra, respectively (shown by the dotted and solid lines in parts a and b of Figure~\ref{fig:amorph_struct}). From this we estimate the ratio of point-sharing to edge-sharing polyhedra is approximately 2:1 in our amorphous structures. We also note that the width of each peak in the RDF is close to 0.5~\AA, indicating that there is significant variation between crystal field environments of each Nb atom.

\section{Machine Learning Details} \label{sec:si_ml}

To explore correlations between local structure and formation of local magnetic moments, we use the random forest classifier implementation in the scikit-learn package of Python. The random forest classifier was trained using the nine stoichiometric amorphous structures, and the six amorphous defect structures (single O vacancy, nearest-neighbour double O vacancies, well-separated double O vacancies, single O interstitial, double O interstitial, and Nb vacancy). Structural features were extracted from nearest neighbor bonding environments for each ion using the CrystalNN, and VoronoiNN packages in Pymatgen~\cite{pymatgen}, which we describe in more detail below.

\begin{figure}[h]
    \centering
    \includegraphics[width=\linewidth]{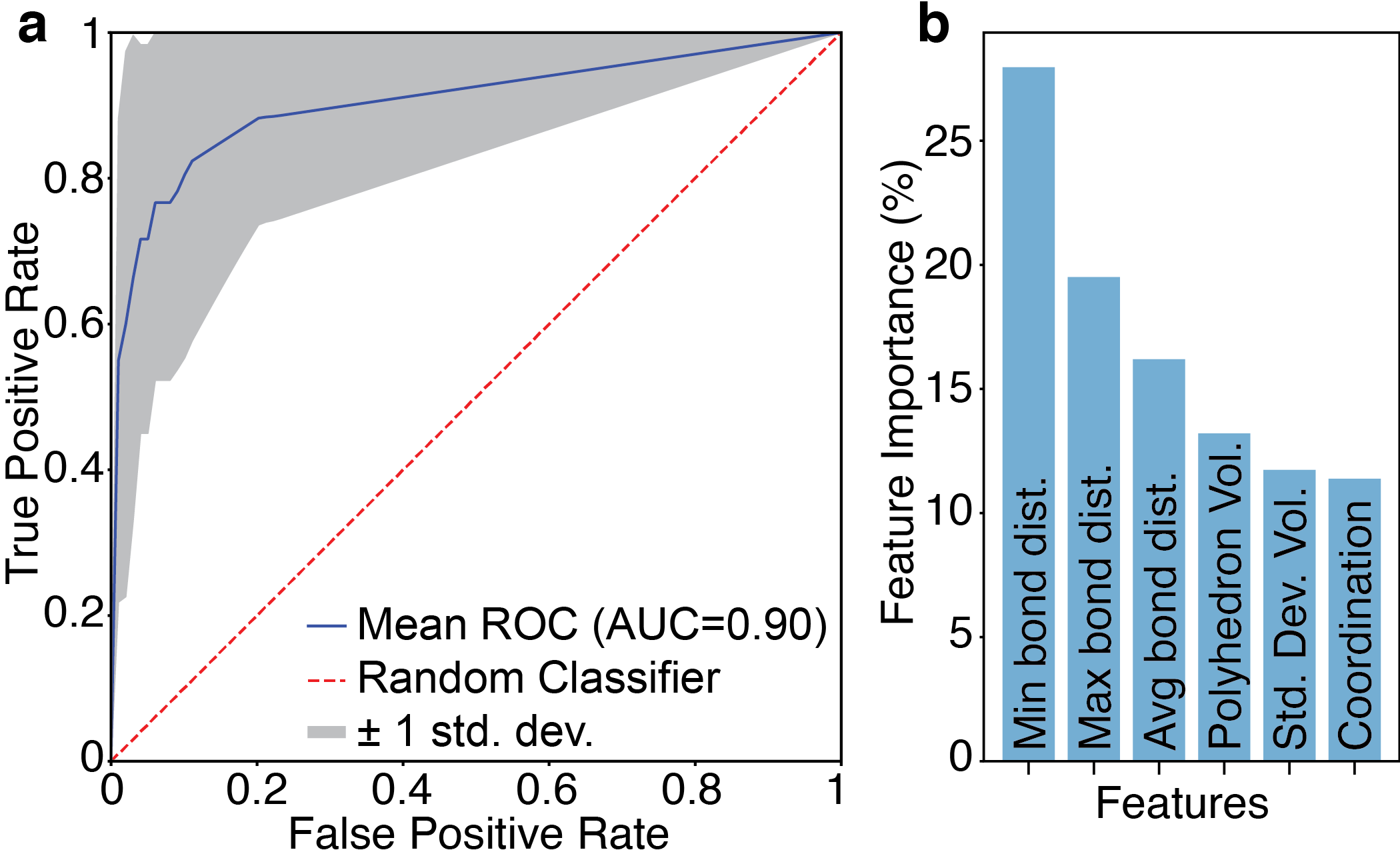}
    \caption{(a) Receiver-operating characteristic curve of a random forest model trained to classify if an ion is magnetic depending on local environment descriptors of the amorphous configurations. (b) Histogram of feature importance (percentage) of a pruned set of descriptors. }
    \label{fig:amorph_rf}
\end{figure}

To create a random forest classifier, we gathered descriptors of local bonding environments for each ion in all amorphous configurations including the set of nearest neighbours, bond lengths, the Voronoi polyhedron describing the local connectivity using the CrystalNN and VoronoiNN packages in Pymatgen\cite{pymatgen}. Each ion within each structure is considered an independent magnetic site, resulting in a dataset with 1576 members. We chose a max tree depth of 20, and  the number of trees in the forest was 200. The threshold for considering a site magnetic in the dataset is 0.05~$\mu_B$.


We initially trained the classifier using all the descriptors above, which were later pruned for co-dependent descriptors. Cross-validation of the random forest was performed using a 5-fold split as implemented in the \textit{KFold} object in scikit-learn with Python\cite{scikit-learn} where each split was used to create a receiver-operating characteristic (ROC) curve. The ROC curve is the comparison of the true positive rate with the false positive rate for a binary classifier where here we define the true positive rate as the fraction of trees that correctly assign the magnetic classification to a test ion.


In Figure~\ref{fig:amorph_rf}a, we plot the ROC curve to demonstrate the accuracy of the random forest classifier. Using the area under the curve (AUC) as a metric for random forest quality, we assessed the impact of different descriptors. We found that only six descriptors were necessary to maintain the AUC value shown in Figure~\ref{fig:amorph_rf}. The most relevant features left after pruning were: (1) minimum, (2) maximum, and (3) mean bond-distance amongst nearest neighbours, (4) Voronoi polyhedron volume, (5) standard deviation of the tetrahedra that comprise the full Voronoi polyhedron, and (6) coordination number. This result suggests that the type of polyhedron, determined via Voronoi tesselation and characterized via the Voronoi indices, does not affect the formation of local magnetic moments. 



In Figure~\ref{fig:amorph_rf}b, we show the feature importances of each of the descriptors used in the classifier, which demonstrates that the bond lengths are most important for determining magnetism, consistent with the tendency for magnetism to form on localized orbitals. This is followed by polyhedron shape/distortion, and coordination. The minimum bond distance is the most important descriptor. For most decision trees in the forest, the first decision was to test if the minimum bond distance between an ion and its neighbours is greater than a threshold value, then it is considered magnetic. If not, then there were many other decisions to be considered. If the bond distances are increased by  strain or negative pressure, for example, this would increase the tendency for local magnetic moments, consistent with our negative pressure results presented in the main text.







\section{Shiba Theory} \label{sec:si_shiba}

The theoretical outline of the microscopic surface impedance is detailed in section \ref{sec:si_surface_impedance}, here we discuss the choice of parameters used. We extract $J$ and $\tau_{s}^{-1}$ from the  \textit{ab-initio} calculations of crystalline Nb$_{2}$O$_{5}$. For a rigid band calculation corresponding to Nb$_{2}$O$_{5.02}$ we find that $J = E^{d} - E_{\text{nm}}^{d} = 45.8 $ meV per u.c; and for Nb$_{2}$O$_{4.98}$ we find that $J = E^{d} - E_{\text{nm}}^{d} = 8.5 $ meV per u.c. Therefore, the ratio between exchange coupling is $\sim 6$ per u.c which gives a lower bound on the input values for our parameterised Shiba Theory. Increasing the O doping concentration tends to increase the exchange coupling, and in the case of O doping concentration of $\sim 0.2$ per u.c  we find that $J_{d}/J_{p} \sim 9$ per u.c.

For the scenario of explicit doping we count the number of magnetic ions with magnetic moment $>0.01 \mu_B$ for varying defect concentrations, as reported in the main text. We find that the number of magnetic O defects created when hole doped is approximately twice that of the number of Nb defects when electron doped, for varying O doping concentrations. As noted in the main text, the final outcome for the magnetic moment is dependent on the size of integration sphere used, which has a direct impact on the choice of bounds we use for the Shiba Theory of surface impedance. However, we comment here that this will not have a significant impact on our parameter selection, due to the sizeable energetic difference between the \textit{p}- and \textit{d}-channel magnetic configurations. Thus, we expect that our estimations on both the magnetic impurity number density and exchange interaction to be reliably calculated from the ab-initio simulation.

\section{Ab-initio Shiba theory for surface impedance} \label{sec:si_surface_impedance}

\subsection{YSR-dissipation}
We consider the low frequency limit ($\omega \sim 0.01$ meV and $|\Delta^{SC}| \sim 1$ meV), where $|\Delta^{SC}|$ is the superconducting gap,  when the current response function can be approximated by:

\begin{equation} \label{eq:q2}
Q_{2}(\omega,k) = Q_{0} \omega \int_{\infty}^{\infty} d\varepsilon \left( -\frac{dn_{0}(\varepsilon)}{d\varepsilon} \bar{Q}_{2}(\varepsilon,k)\right).
\end{equation}
Here, $n_{0} = 1 / (\exp(\varepsilon/T) +1) $ is the Fermi-Dirac distribution, and $\bar{Q}_{2}(\varepsilon,k)$is given by 
\begin{equation}
\begin{split}
    \bar{Q}_{2}(\varepsilon,k) &= \frac{3}{2} ( [\mathcal{F}(\varepsilon)]^{2} \langle \bar{jj}\rangle^{RR}(\varepsilon,0,k) \\
    &+ [\mathcal{F}^{*}(\varepsilon)]^{2} \langle \bar{jj}\rangle^{AA}(\varepsilon,0,k) \\
    &+ [1 + |\mathcal{G}(\varepsilon)|^{2} + |\mathcal{F}(\varepsilon)|^{2}]\langle \bar{jj}\rangle^{RA}(\varepsilon,0,k)),
\end{split}
\label{eq:q2b}
\end{equation}

where $\mathcal{F}(\varepsilon)$ is the quasi-classical anomalous Green's function and $\langle \bar{jj}\rangle^{i}$ are the constituent retarded and advanced response functions. For $k=0$ we analyse $\bar{Q}_{2}(\varepsilon,k)$ which describes the dissipative part of the current response function $Q(\omega,k)$ originating from quasiparticles within an energy interval $\varepsilon$ around the Fermi level.

\begin{figure}[h]
    \centering
    \includegraphics[width=0.8\linewidth]{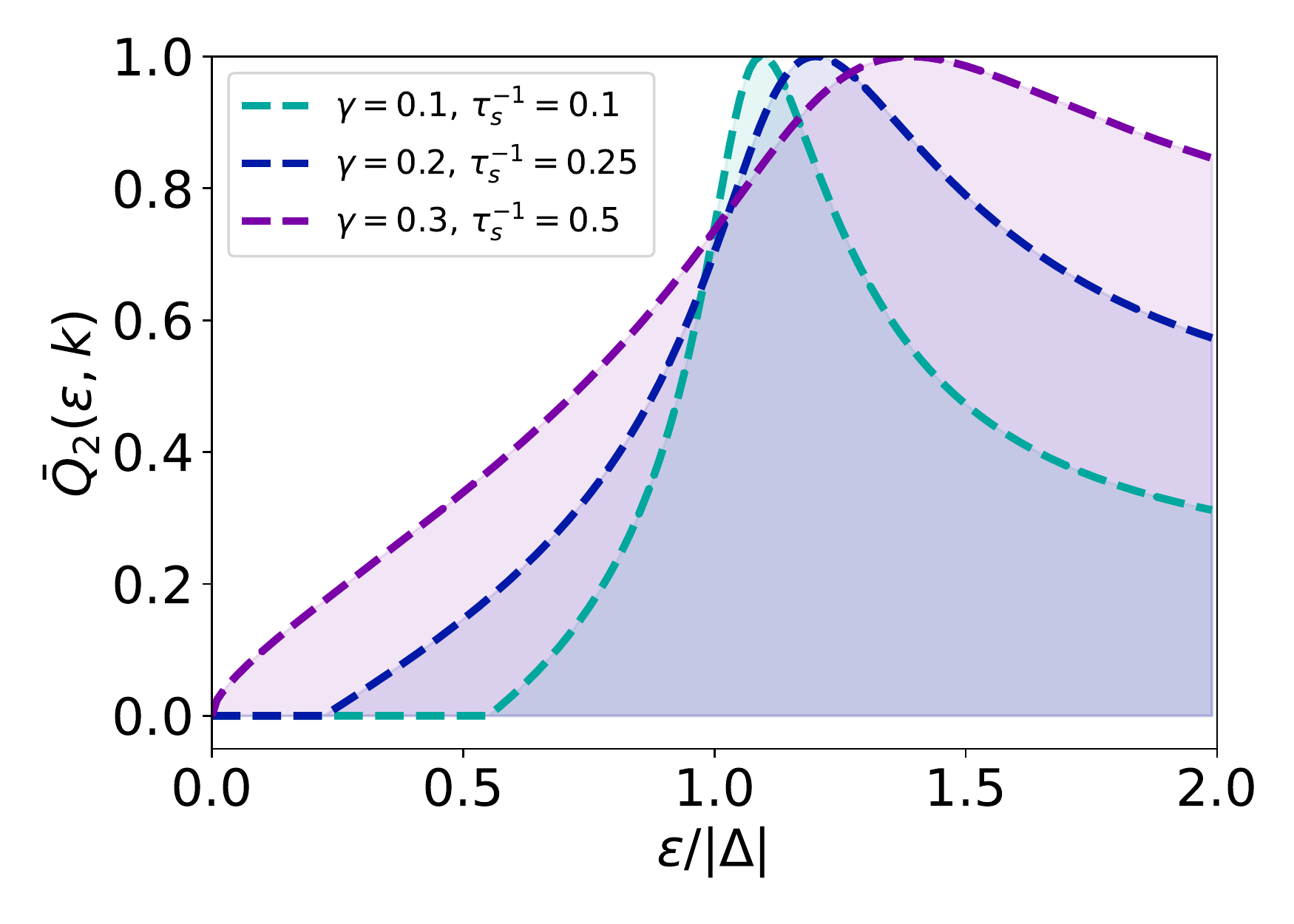}
    \caption{The real part of the current dissipation function for \textit{p}-channel losses in the Nb-oxide for increasing exchange coupling and magnetic impurity density. $\tau_s$ is the scattering time on magnetic impurities and $\gamma$ is the Shiba exchange coupling constant. For sufficiently large impurity density and exchange coupling the dissipation function is non-zero at the Fermi level, thus contributing to the impedance indicating a non-negligible association of the \textit{p}-channel magnetism to residual losses in the Nb-films. }
    \label{fig:SI_shiba}
\end{figure}

For \textit{d}-channel magnetism it is found out-of-the-box that there are significant decoherence losses as evidenced in the current response functions and surface impedance in the low frequency limit. For \textit{p}-channel magnetism we demonstrate that it is possible to induce a residual resistance for \textit{p}-type decoherence losses, but it requires a much higher density of magnetic impurities than those reported from ab-initio theory. In Fig.~\ref{fig:SI_shiba} we demonstrate the cooperative behaviour between exchange coupling and impurity density. For a critical values of $\gamma=0.3$ and $\tau_{s}^{-1}=0.1$ we find that scenario in which \textit{p}-channel magnetism can induce non-negligible current dissipation. 

In the low frequency limit the surface resistance $R(\omega)$ can be approximated as:
\begin{equation}
    Z(\omega) = \frac{32 \pi \omega}{c^{4}} \int_{0}^{\infty} dk \frac{Q_{2}(\omega,k)}{[k^{2} + 4\pi Q_{1}(k)/c^2]^{2}}. 
\end{equation}
The surface resistance $R(\omega)$, determined by $Z(\omega) = R(\omega) + i X(\omega)$, is quadratic in frequency, i.e $R(\omega) \propto \omega^2$. 

\bibliography{refs}